\documentclass{jaa}

\usepackage{graphicx}
\usepackage{amsmath}
\usepackage{mathtools}
\usepackage{tablefootnote} 
\usepackage{url} 

\usepackage[dvipsnames]{xcolor} 

\newcommand\arcdeg{\mbox{$^\circ$}}%
\newcommand\arcmin{\mbox{$^\prime$}}%
\newcommand\arcsec{\mbox{$^{\prime\prime}$}}%

\newcommand{\js}{} 

\newcommand{\etal}{{\it et~al.~\/}}   
\newcommand{\dodoi}[1]{\textcolor{blue}{{https://doi.org/#1}}}
\newcommand{\cc}{\textcolor{blue}}   

\newcommand\araa{\js{ARA\&A}}
\newcommand\apj{\js{ApJ}}
\newcommand\apjl{\js{ApJL}}     
\newcommand\apjs{\js{ApJS}}
\newcommand\lrsp{\js{LRSP}}
\newcommand\ao{\js{ApOpt}}
\newcommand\aap{\js{A\&A}}
\newcommand\aapr{\js{A\&A~Rv}}
\newcommand\icarus{\js{Icarus}}
\newcommand\mnras{\js{MNRAS}}
\newcommand\prl{\js{PhRvL}}
\newcommand\pasj{\js{PASJ}}
\newcommand\solphys{\js{SoPh}}
\newcommand\ssr{\js{SSRv}}
\newcommand\zap{\js{ZA}}
\newcommand\nat{\js{Nature}}
\newcommand\grl{\js{GeoRL}}
\newcommand\jgr{\js{JGR}}
\newcommand\planss{\js{Planet.~Space~Sci.}}
\newcommand\jastp{\js{JASTP}}

\begin{document}\sloppy

\title{Propagation of Coronal Mass Ejections from the Sun to Earth}


\author{Wageesh Mishra\textsuperscript{1*} and Luca Teriaca\textsuperscript{2}}
\affilOne{\textsuperscript{1}Indian Institute of Astrophysics, II Block, Koramangala, Bengaluru 560034, India \\} 
\affilTwo{\textsuperscript{2}Max Planck Institute for Solar System Research, G\"ottingen 37077, Germany \\}


\twocolumn[{

\maketitle

\corres{wageesh.mishra@iiap.res.in}

\msinfo{***}{***}


\begin{abstract}
Coronal Mass Ejections (CMEs), as they can inject a large amounts of mass and magnetic flux into the interplanetary space, are the primary source of space weather phenomena on the Earth. The present review first briefly introduces the solar surface signatures of the origins of CMEs and then focuses on the attempts to understand the kinematic evolution of CMEs from the Sun to the Earth. CMEs have been observed in the solar corona in white-light from a series of space missions over the last five decades. In particular, LASCO/\textit{SOHO}  has provided almost continuous coverage of CMEs for more than two solar cycles until today. However, the observations from LASCO suffered from projection effects and limited field of view (within 30 R$_\odot$ from the Sun). The launch in 2006 of the twin \textit{STEREO} spacecraft made possible multiple viewpoints imaging observations, which  enabled us to assess the projection effects on CMEs. Moreover, heliospheric imagers (HIs) onboard \textit{STEREO}  continuously observed the large and unexplored distance gap between the Sun and Earth. Finally, the Earth-directed CMEs that before have been routinely identified only near the Earth at 1 AU in \textit{in situ} observations from \textit{ACE} and \textit{WIND}, could  also be identified at longitudes away from the Sun-Earth line using the \textit{in situ} instruments onboard \textit{STEREO}. We describe the key signatures for the identification of CMEs using \textit{in situ} observations. Our review presents the frequently used methods for estimation of the kinematics of CMEs and their arrival time at 1 AU using primarily \textit{SOHO} and \textit{STEREO} observations. We emphasize the need of deriving the three-dimensional (3D) properties of Earth-directed CMEs from the locations away from the Sun-Earth line. The results improving the CME arrival time prediction at Earth and the open issues holding back progress are also discussed. Finally, we summarize the importance of heliospheric imaging and discuss the path forward to achieve improved space weather forecasting.
\end{abstract}


}]


\doinum{000.00}
\artcitid{\#\#\#\#}
\volnum{000}
\year{0000}
\pgrange{1--}
\setcounter{page}{1}
\lp{1}

\section{Introduction}
\label{Intro}


The extremely hot, tenuous and outermost atmosphere of the sun is called the solar corona. This extends to several millions of kilometres above the visible surface of the Sun (i.e., solar photosphere) and is much fainter than the photosphere. The solar corona is naturally seen in visible light only during a total solar eclipse when the moon shadows the bright photosphere.  The solar corona is also observable with an instrument called coronagraph which was introduced in 1931 by the French astronomer Bernard Lyot (\cc{Lyot 1939}). A coronagraph creates an artificial eclipse by selectively blocking out the photospheric light from the disk of the Sun so to observe the corona.

It is now understood that the solar corona releases a constant out-stream of energized charged particles which is called solar wind (\cc{Biermann 1951}; \cc{Parker 1958}). The solar wind fills the interplanetary space and its existence was first confirmed by direct observations from spacecraft Luna 1 (\cc{Gringauz \etal 1960}). In addition to ubiquitous solar wind, the solar corona frequently expels large-scale magnetized plasma structures into the heliosphere. Such episodic expulsions of plasma from the Sun are called Coronal Mass Ejections  (CMEs). The earliest observation of a CME probably dates back to the eclipse of 1860 as clearly seen in a drawing recorded by G. Temple. Some definite inferences for CMEs from the Sun were made before their formal detections (\cc{Chapman \& Ferraro 1931}; \cc{Eddy 1974}). However, CMEs were first detected in 1971 by a coronagraph onboard NASA’s seventh \textit{Orbiting Solar Observatory (OSO-7)} satellite (\cc{Tousey 1973}). The name CME was initially coined for a feature which shows an observable change in coronal structure that occurs on a time scale of few minutes to several hours and involves the appearance (and outward motion) of a new, discrete, bright, white-light feature in the coronagraphic field of view (FOV) 
(\cc{Hundhausen \etal 1984}).

\begin{figure}[!htb]
\centering
\includegraphics[scale=0.4]{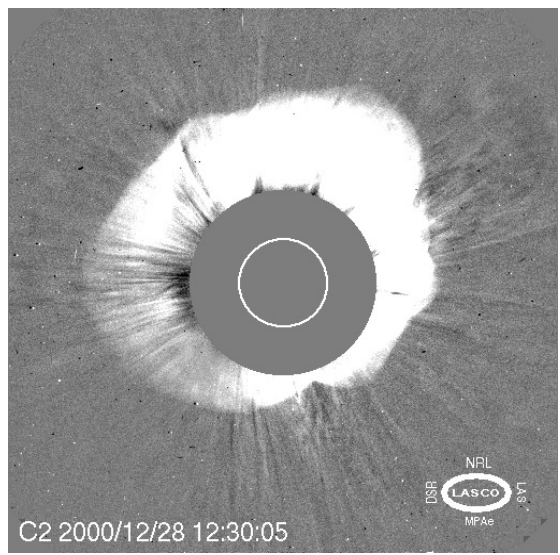}
\caption[A ``halo'' CME observed by LASCO-C2 coronagraph on \textit{SOHO}]{An image of a ``halo'' CME observed by LASCO-C2 coronagraph onboard \textit{SOHO}. The CME was launched from the Sun on 2000 December 28. The white circle in the center is the size and location of the solar disk, which is obscured by the coronagraph's occulter, covering up to 1.7 \textit{R$_\odot$}. (\cc{\textit{Image credit: \url{http://lasco-www.nrl.navy.mil}})}}
\label{haloCME}
\end{figure}

The observations of CMEs have been made using white-light coronagraphs, interplanetary scintillation measurements, and \textit{in situ} observations. The coronagraphs record a two-dimensional (2D) image of a three-dimensional (3D) CME projected onto the plane of the sky. Therefore, the morphology of CME in coronagraphic observations depends on the location of the observing instruments (e.g., coronagraphs) and the launch direction of CME from the Sun. The CMEs launched from the Sun toward or away from the Earth, when observed by the near-Earth coronagraphs will appear as `halos' surrounding the occulting disk of coronagraphs (\cc{Howard \etal 1982}). Such a CME is called a ``halo'' CME  (Figure~\cc{\ref{haloCME}}). An example of coronagraph observing from near Earth is \textit{ Large Angle Spectrometric COronagraph } (LASCO) onboard \textit{SOlar and Heliospheric Observatory (SOHO)} located at the L1 point of the Sun-Earth system. A CME having 360$\arcdeg$ apparent angular width is called ``full halo'' and with apparent angular width greater than 120$\arcdeg$ but less than 360$\arcdeg$ is called as ``partial halo''. Such a nomenclature of a CME is restricted by its viewing perspective. The observations of solar activity on the solar disk, associated with CME, are necessary to help distinguish whether a halo CME was launched from the front or backside of the Sun relative to the observer. It is important to note that among all the CMEs, only about 10\% are partial halo type (i.e. width greater than 120$\arcdeg$)  and about 4\% are full halo type (\cc{Webb \etal 2000}).

The CMEs observed as front-side halo are important as they are the key link between solar eruptions and major space weather phenomena. The term space weather refers to conditions in the space between the Sun and Earth (e.g., in the solar wind, Earth's magnetosphere, ionosphere, and thermosphere) that can influence the performance and reliability of space-borne and ground-based technological systems and can endanger human life or health. The majority of geomagnetic storms of solar cycles 23 and 24 are known to be caused by halo CMEs, confirming the importance of the source location of CMEs (\cc{Gopalswamy 2010}; \cc{Lawrance \etal 2020}). The source regions of front-side halo CMEs can be studied in greater detail with instruments capable of imaging the structures at the base of the corona. The example of such instruments are \textit{Extreme-ultraviolet Imaging Telescope} (EIT) onboard \textit{SOHO}, \textit{Atmospheric Imaging Assembly} (AIA) onboard \textit{Solar Dynamics Observatory (SDO)} and \textit{Extreme-Ultraviolet Imager} (EUVI) as a part of \textit{Sun Earth Connection Coronal and Heliospheric Investigation} (SECCHI) package onboard \textit{Solar TErrestrial RElations Observatory (STEREO)} (\cc{Delaboudini{\`e}re \etal 1995}; \cc{Lemen \etal 2012}; \cc{Howard \etal 2008}). If such CMEs do not get a large deflection during their interplanetary propagation, they are expected to be sampled at observer's location by \textit{in-situ} spacecraft (\cc{Webb \etal 2000}). It is important to note that CMEs are the 3D structure, therefore single-point imaging observations would suffer from the unavoidable projection effect (\cc{Burkepile \etal 2004}). In the case of a halo CME, the projection effects are considerably large and the measured speed of a CME is underestimated while its angular width is overestimated (\cc{Xie \etal 2004}). The CME's initial speed, angular width, direction, and background solar wind are known to govern the transit time of the CME from the Sun to 1 AU (\cc{Gopalswamy \etal 2000a}; \cc{M{\"o}stl \& Davies 2013}). It is shown that even CMEs of equal speeds but different geometry and propagation direction can take quite different transit times to reach Earth. Therefore, the kinematic and geometric parameters of halo CMEs need further corrections for accurate forecasting of their arrival time (\cc{Shen \etal 2014}). In addition to forecasting purpose, the projection effects on halo CMEs also impose limitations on our understanding of physical characteristics of CMEs.

\begin{figure}[!htb]
\centering
\includegraphics[trim = 10mm 76mm 165mm 34mm, clip]{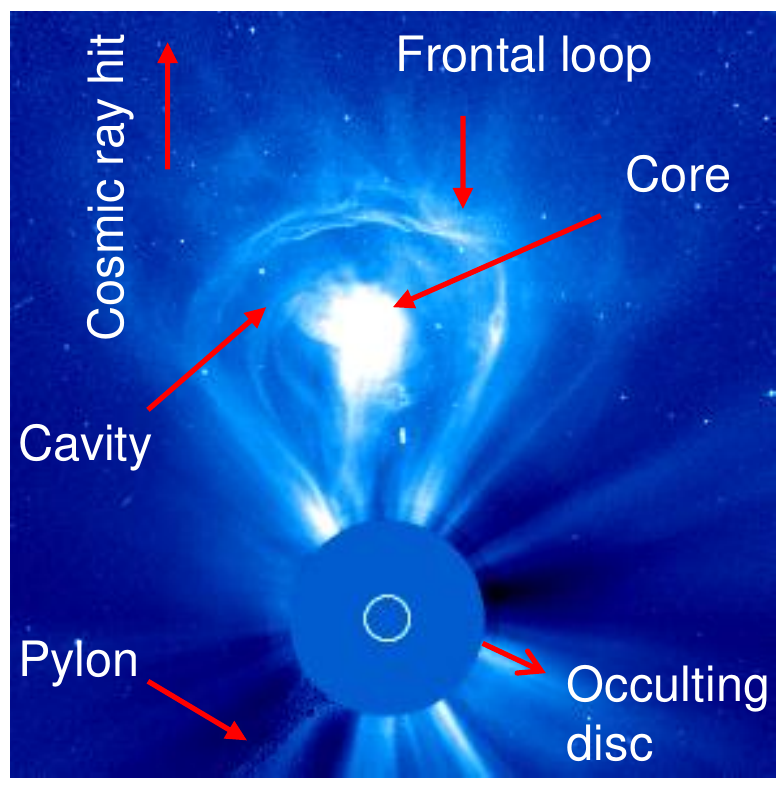}
\caption[A classical CME with 3-part structure]{A classical 3-part CME seen in the LASCO-C3 field of view on 2000 February 27 at 07:42 UT, showing a bright frontal loop surrounding a dark cavity with a bright core. (\cc{\textit{Image credit: \url{http://soho.nascom.nasa.gov}})}}
\label{threepartCME}
\end{figure}

Some CMEs observed near the Sun often appear as a ``three-part'' structure comprising of an outer bright frontal loop (i.e. leading edge), and a darker underlying cavity within which is embedded a brighter core as shown in Figure~\cc{\ref{threepartCME}} (\cc{Illing \& Hundhausen 1985}). The front may contain swept-up material by erupting flux ropes or the presence of pre-existing material in the overlying fields (\cc{Illing \& Hundhausen 1985}; \cc{Riley \etal 2008}). The cavity is a region of lower plasma density but probably higher magnetic field strength, i.e., a manifestation of a driving flux rope (\cc{Forsyth \etal 2006}). The brightest component of the three-part structure, i.e., the core of the CME can often be identified as prominence (i.e., filament) material based on their visibility in chromospheric emission lines  (\cc{Bothmer \& Schwenn 1998}; \cc{Schmieder \etal 2002}).  Contrary to an established perspective held for several decades, recently it has been shown that bright cores can be observed in many CMEs which are not associated with filament eruptions in any way (\cc{Howard \etal 2017}; \cc{Song \etal 2017}). Moreover, they found that in some cases where CMEs were associated with filament eruptions, the bright cores neither geometrically resemble eruptive filaments nor exhibit H$\alpha$ emission as expected from cool filament materials in the coronagraphic field of view. Based on this, \cc{Howard \etal (2017)} suggested that the bright core within the cavity could be an optical illusion produced by the geometrical projection of a twisted 3D flux rope into a 2D plane or it can appear due to the natural evolution of an erupting flux rope (\cc{Howard \etal 2017}).

It is noted that the frequency of occurrence of CMEs around solar maximum is $\approx$ 5 per day and at solar minimum is $\approx$ 4 per week (\cc{St. Cyr \etal 2000}; \cc{Webb \& Howard 2012}).  CMEs having a three-part structure are only about 30\% of all the CMEs from the Sun, yet this is considered as the ``standard CME'' configuration in observational and theoretical studies (\cc{Gopalswamy 2004}; \cc{Gopalswamy 2006a}). Despite the common association of CMEs with eruptive filaments and flux ropes, surprisingly only about 4\% of the Earth-arriving ICMEs show the signatures of filaments and only about 35\% of ICMEs show the signatures of flux ropes in \textit{in-situ} observations at 1 AU (\cc{Lepri \& Zurbuchen 2010}; \cc{Richardson \& Cane 2010}).  The absence of flux rope in some ICMEs is understood in term of geometric selection effect (\cc{Kilpua \etal 2011}; \cc{Song \etal 2020}), but the rarely observed filaments at large distances from the Sun pose a question if they survive at all beyond a few solar radii from the Sun. There are case studies that have shown that soon after the launch of a filament from the Sun, it may get fragmented into magnetic Rayleigh–Taylor (MRT) unstable plasma segments and fall back into the solar atmosphere (\cc{Innes \etal 2012}; \cc{Mishra \etal 2018a}; \cc{Mishra \etal 2018b}). \cc{Joshi \etal (2013)} have shown a case study where the core of a CME associated with an asymmetric filament eruption exhibited downfall of its plasma which they explained using a self-consistent model of a magnetic flux rope. Thus, the draining of filament plasma can be partly responsible for their absence in coronagraphic and \textit{in situ} observations. Also, the ionization of the filament material can take place during its evolution away from the Sun (\cc{Howard 2015}), and this can make them spread out across their respective field lines and become indistinguishable from the material making up the surrounding CME.

It is known that not all CMEs appear to have a very large angular width in coronagraphic images, in fact, some CMEs appear as narrow jets. However, it should be noted that wide CMEs are not necessarily very global but rather may have a propagation direction along the Sun-observer line, and so they appear large by perspective as noted for the so-called halo CMEs. CME's are classified as narrow when they have an apparent angular width less than 20$^{\circ}$ and they are a small subset of all CMEs (\cc{Yashiro \etal 2003}). The average width of normal three-part structure CMEs has been reported to range from 50$^{\circ}$ to 70$^{\circ}$ depending on the inclusion of partial halos, full halos, and different phase of a solar cycle (\cc{St. Cyr \etal 2000}; \cc{Webb \& Howard 2012}). Based on the LASCO CMEs in the CDAW catalog (\cc{Gopalswamy \etal 2010}),  narrow CMEs are found to be only about 12\% and 22\% of the total number of CMEs during the minimum and maximum of solar cycle 23, respectively (\cc{Yashiro \etal 2003}). According to \cc{Gilbert \etal (2001)}, the average speeds of narrow CMEs are similar to that of normal CMEs. The speeds of narrow CMEs near the Sun range from few km s$^{-1}$ to 1150 km s$^{-1}$ but for the normal CMEs it can range from few km s$^{-1}$ to 3000 km s$^{-1}$ (\cc{St. Cyr \etal 2000}). On the other hand, \cc{Yashiro \etal (2003)} find that narrow CMEs tend to be faster than normal CMEs during solar maximum. The average mass of a narrow CME is less than about 10\% of the mass of a normal CME which is $\sim$1.5 $\times$10$^{12}$ kg. It has been found that narrow CMEs are the outward extensions of EUV jets and they probably have different acceleration mechanisms than normal CMEs (\cc{Wang \etal 1998}). Recent studies have focused on investigating the triggering mechanism and kinematics of jet-like CMEs (\cc{Solanki \etal 2019, 2020}). Also, studies have reported the simultaneous launch of jet-like and bubble-like CMEs to investigate their eruption mechanisms (\cc{Shen \etal 2012a}; \cc{Duan \etal 2019}).

The triggering and driving mechanisms of CMEs have been the subject of extensive research aimed at developing CME initiation models constrained by observations (\cc{Chen 2011}). The launch of CMEs requires that the magnetic field lines must be opened by some processes to allow the plasma to escape from the Sun. The onset of CMEs has been associated with many solar disk phenomena such as flares (\cc{Feynman \& Hundhausen 1994}), prominence eruptions (\cc{Hundhausen 1999}), coronal dimming (\cc{Sterling \& Hudson 1997}), arcade formation (\cc{Hanaoka \etal 1994}). In fact, it has been observed that the CMEs often show spatial and temporal relation with solar flares, eruptive prominences (\cc{Munro \etal 1979}; \cc{Webb \& Hundhausen 1987}; \cc{Gopalswamy \etal 2003b}) and with helmet streamer blowouts. Solar flares are observed as localized sudden brightening on the Sun across all wavelengths at the time scale of minutes (\cc{Aschwanden 2002}; \cc{Benz 2008}) and historically they were considered to be the drivers for CMEs and interplanetary shocks. Many strong CMEs are associated with intense flares but most of the flares are “confined” or “compact” and occur independently of CMEs, and thus there is no one-to-one relationship between flares and CMEs (\cc{Yashiro \etal 2008}).

Based on several studies in the last two decades, it seems that CMEs and flares are part of a single magnetically-driven phenomenon which creates a larger net energy reservoir available for both phenomena (\cc{Compagnino \etal 2017}). Therefore, a unified standard flare model known as Flux Cancellation or Catastrophe model has been developed and refined over the last few decades (\cc{Forbes \& Isenberg 1991}; \cc{Priest \& Forbes 2002}). Another model called Breakout model has been developed to describe the association of CMEs with flares (\cc{Antiochos \etal 1991}). Therefore, it is evident that although CMEs and flares may not be causally related, they both seem to be involved with the reconfiguration of complex magnetic field lines within the corona caused by the same underlying physical processes, e.g., magnetic reconnection (\cc{Priest \& Forbes 2002}; \cc{Compagnino \etal 2017}).

Furthermore, the eruption of prominences is also often associated with CMEs, with the erupted material forming their bright cores. The prominences are caused by the formation of flux ropes low in the sheared magnetic structure in the corona but they are about one hundred times cooler and denser than the corona. It is established that prominences appear as bright features at the limb but appear darker than their background on the solar disk where they are called filaments. It is now suggested that perturbations (i.e., precursor activities) in coronal magnetic fields forming a CME begin well before any observed associated surface activity such as flares or erupting prominences (\cc{Gopalswamy \etal 2006}). Some of the CMEs are known to appear from the blowout of a helmet streamer due to dynamical evolution of arcades, flux emergence, or shearing of magnetic field lines (\cc{Vourlidas \etal 2002a}, \cc{Gopalswamy  2006a}). A streamer is a dense structure containing closed and open fields which are observed by coronagraphs above the solar limb.

In the attempt to establish the association between CMEs observed in coronagraphs and their signatures at the solar surface, it has been noted that some CMEs do not have easily identifiable signatures (such as coronal dimming, coronal wave, filament eruption, flare, post-eruptive arcade) to locate their source regions on the Sun (\cc{Ma \etal 2010}; \cc{Vourlidas \etal 2011}). Such CMEs are called the ``problem or stealth CMEs'' \cc{Robbrecht \etal 2009}. On comparing CMEs with and without low coronal signatures it is found that stealth CMEs are slow, typically from 100 km s$^{-1}$ to 300 km s$^{-1}$ having gradual acceleration and their source regions are usually located in the quiet Sun in proximity with coronal holes rather than active regions (\cc{Ma \etal 2010}; \cc{Nitta \& Mulligan 2017}). Some stealth CMEs are found to be narrow but they can be wide enough to show the typical three-part structure of the CME. It was suggested by \cc{D'Huys \etal (2014)} that the physical process such as reconnection that enables the stealth CMEs probably occurs at higher altitude. They found that in most of the cases a stealth CME was preceded by another nearby CME which might have destabilized the coronal magnetic field making a path for the stealth CME. The modeling of stealth CMEs by \cc{Lynch \etal (2016)} confirmed the results of \cc{Howard \& Harrison (2013)} that there is no fundamental difference between stealth CMEs and most of the slow streamer blowout CMEs. The initiation mechanism and geoeffectiveness of stealth CMEs associated with the eruption of coronal plasma channel and jet-like structures have also been studied recently (\cc{Mishra \& Srivastava, 2019}).

The important lesson from earlier studies on the origin of CMEs is that although the physical processes making the eruption of CMEs differ in different models, the overall idea is essentially the same: a magnetic field configuration initially kept in equilibrium needs to be disturbed somehow to make the system erupt.  One possibility is that the initial configuration may have an underlying sheared magnetic field (often called core) held down by an overlying field and the equilibrium can be disrupted by the magnetic reconnection between the sheared magnetic core and the overlying field. This can lead to the reconfiguration of magnetic field lines causing eruptions beyond the overlying fields (\cc{Antiochos \etal 1991}). There also exists a scenario of accumulating twist in the magnetic core leading to kink instability, which can push aside the overlying field and make eruption possible (\cc{T{\"o}r{\"o}k \& Kliem 2005}). Once the eruption of a CME has taken place, then the remaining magnetic field eventually closes, probably via some form of large-scale magnetic reconnection. It is noted that despite the development in understanding the origin of CMEs, the models are inadequate to completely match observations of an evolving CME under pressure, magnetic and gravitational forces (\cc{Gopalswamy 2004}; \cc{Webb \& Howard 2012}).

CMEs can lead to disturbances in the heliosphere, from their birth-place in the corona up to several AU distances away from the Sun, e.g., interplanetary shocks, radio bursts, intense geomagnetic storm, large solar energetic particles (SEPs) events and Forbush decreases (FDs) (\cc{Kahler \etal 1978}; \cc{Gosling 1993}; \cc{Wang \etal 2000}; \cc{Gopalswamy \etal 2000b}; \cc{Gopalswamy 2006b}; \cc{Richardson \& Cane 2010}; \cc{Wiedenbeck \etal 2020}).  It is shown that SEP events are associated with fast and wide CMEs near the Sun (\cc{Gopalswamy \etal 2003a}). The accelerated electrons by CME-driven shocks can produce Type II radio bursts that appear as slowly drifting features in radio dynamic spectra (\cc{Gopalswamy \etal 2013}). The distance of CME from the Sun at the time of onset of Type II bursts is the height where the CME becomes super-Alfv\'enic  to drive fast mode MHD shocks. The height of shock formation is important in understanding the SEPs and their charge states. The studies on shock formation height suggest that shocks related to metric and decameter-hectometric (DH) type II bursts form at heights smaller and larger than 2 R$_\odot$ from the center of the Sun, respectively (\cc{Ramesh \etal 2012}; \cc{Gopalswamy \etal 2013}). Studies using extreme ultra-violet and white-light imaging observations of CMEs have suggested that the type II bursts can originate from anywhere on the shock front (i.e., at the nose or flanks) depending on which location is favorable for electron acceleration. The pre- and post-shock parameters of coronal plasma were studied by \cc{Bemporad \& Mancuso (2010)} and they found an increase in plasma temperature and magnetic field across the shock. The effects of shock compression have also been noted in the \textit{in situ} observations at 1 AU in the scenario where the following shocks penetrated through preceding ICMEs (\cc{Harrison \etal 2012}; \cc{Liu \etal 2012}; \cc{Mishra \& Srivastava 2014}).

CMEs and their driven shocks are found to interact with the atmosphere and magnetosphere of planets leading to severe space weather activity (\cc{Wang \etal 2003}; \cc{Schwenn 2006}; \cc{Baker 2009}; \cc{Mishra \etal 2015a}; \cc{Luhmann \etal 2020}). A typical example of a space weather event is the geomagnetic storm in which a major disturbance of Earth's magnetosphere takes place due to the efficient transfer of energy from the solar wind into the space environment surrounding Earth. The effect of CMEs on a planet is governed by the magnetic nature of the planet. The Earth has a magnetic field and hence Earth-arriving ICME structures having strong southward magnetic field component (\textit{B$_{z}$}), interact with the Earth’s magnetosphere at the day-side magnetopause. In this interaction, solar wind energy is transferred to the magnetosphere, primarily via magnetic reconnection that produces non-recurrent geomagnetic storms (\cc{Dungey 1961};  \cc{Tsurutani \etal 1988}; \cc{Gonzalez \etal 1994};  \cc{Baker 2009}). It has been shown that 83\% intense geomagnetic storms are due to CMEs (\cc{Zhang \etal 2007}). Few of the intense storms may occur because of corotating interaction regions (CIRs). CIRs form when the fast speed solar wind overtakes the slow speed solar wind ahead and leads to the formation of an interface region that has increased temperature, density, and magnetic field. The arrival time of interacting CMEs and their geomagnetic consequences have also been studied extensively (\cc{Farrugia \etal 2006}; \cc{Lugaz \& Farrugia 2014}; \cc{Liu \etal 2014}; \cc{Mishra \etal 2015a}; \cc{Lugaz \etal 2017}). Thus, from a space weather perspective, it is important to estimate the arrival time and transit speeds of CMEs as well as orientation of magnetic field in the CMEs near the Earth well in advance to predict the severity of these events (\cc{Gonzalez \etal 1989}; \cc{Srivastava \& Venkatakrishnan 2002}; \cc{Vourlidas \etal 2019}).

The Earth-arriving CME-driven shock compresses the day-side Earth's magnetosphere and causes the storm sudden commencement (SSC) (\cc{Chao \& Lepping 1974}). The horizontal component of Earth's magnetic field, which can be measured by ground-based magnetometers, is found to be increased during SSC (\cc{Dessler \etal 1960}; \cc{Tsunomura 1998}). Furthermore, the sheath region lying between the shock and flux rope get compressed and may also have negative B$_{z}$. It is also well proven that CMEs are responsible for the periodic 11-year variation in the galactic cosmic rays (GCRs) intensity (\cc{Cane 2000}). Moreover, CMEs are found to be responsible for Forbush decreases (FDs) (\cc{Forbush 1937}). Non-recurrent FDs are defined as a sudden shorter-term decrease of the recorded intensity of GCRs when a CME passes Earth. In FDs, the depression in the intensity of GCRs by 3\% to 20\% typically lasts for minutes to hours, while its recovery to normal level takes place in several days. FDs are due to exclusion of GCRs because of their inability to diffuse ``across'' the relatively strong and ordered IMF in the vicinity of interplanetary shock, in the sheath and/or flux-rope region of the CME. The FDs have also been the focus of many studies to examine a possible connection between the GCR flux and Earth’s climate via modulation of cloud cover (\cc{Lam \& Rodger 2002}; \cc{Laken \etal 2009}). The FDs are routinely measured on the surface of Earth using neutron monitors and can be used to detect the arrival of CMEs and their speeds (\cc{Dumbovi{\'c} \etal 2018}).

Keeping the Sun-Earth connection in mind, several studies have been undertaken in the last decades, before and after the launch of \textit{STEREO}, to understand the propagation of CMEs and estimate their arrival times at Earth. The most recent reviews on ICMEs and their arrival times are by \cc{Kilpua \etal 2017}, \cc{Vourlidas \etal 2019}, \cc{Luhmann \etal 2020}, \cc{Temmer \etal 2021} and \cc{Zhang \etal 2021}. Although much progress has been made in this direction, yet the accurate prediction of arrival times of CMEs remains difficult even today. In this review, we highlight several important earlier studies carried out to reach our present understanding of CMEs propagation. We also discuss open questions holding us back from progressing and the path forward for improving the accuracy in CME forecasting in the near future.

\section{Studies on CME Propagation Before \textit{STEREO} era}
\label{prest}

Although our review is focused on heliospheric propagation of CMEs, we would briefly mention a few of the studies on the origin of CMEs. There is a vast literature on the photosheric and coronal properties of source active regions that produces CMEs (\cc{Cliver \& Hudson 2002}; \cc{Kahler 2006}; \cc{Georgoulis \etal 2019}, and references therein). The initiation and early development of CMEs have been studied since the pioneering work on EUV waves by \cc{Dere \etal (1997)} using the observations of the \textit{Extreme-ultraviolet Imaging Telescope} (EIT) onboard \textit{SOHO}. Recently, the availability of high resolution observations from \textit{Extreme UltraViolet Imager} (EUVI) onboard \textit{STEREO} and the \textit{Atmospheric Imaging Assembly} (AIA) onboard \textit{SDO} have further helped in exploring the solar surface signatures of CMEs (\cc{Vr{\v s}nak \& Cliver 2008}; \cc{Liu \& Ofman 2014}, and references therein).  Furthermore, the densities, temperatures, ionization states, and Doppler velocities of CMEs have been studied using EUV spectral observations from the \textit{UltraViolet Coronagraph Spectrometer} (UVCS), \textit{Coronal Diagnostic Spectrometer} (CDS), and \textit{Solar Ultraviolet Measurements of Emitted Radiation}  (SUMER) instruments onboard \textit{SOHO} and the \textit{Solar Optical Telescope} (SOT), and the \textit{Extreme Ultraviolet Imaging Spectrometer} (EIS)  instruments on Hinode (\cc{Raymond 2002}; \cc{Kohl \etal 2006}; \cc{Landi \etal 2010}).

It is also noted that there have been a plethora of multi-wavelength studies on associating CMEs origins with their signatures on the Sun such as streamers blowouts, solar flares, erupting prominences/filaments, coronal dimming, arcade formations,  and coronal waves (\cc{Tripathi \etal 2004}; \cc{Burkepile \etal 2004};  \cc{Benz 2008}). These signatures of CMEs origin are primarily observed in wavelengths capable of imaging different layers of the solar atmosphere at varying temperatures and also plasma material of different ionization states. This is unlike observations of CMEs by visible light coronagraphs and heliospheric imagers which observe the angular dependent brightness of Thomson-scattered white-light from the free electrons in CMEs. Importantly, the white-light observations often sample the CMEs at heights different than the height where the signatures of CMEs initiations are observed (\cc{Gopalswamy 2004}; \cc{Webb \& Howard 2012}). Therefore, it is difficult to make a direct association of CMEs and their associated phenomena at the solar surface. In the following, we will focus on the white-light and \textit{in situ} observations of CMEs.

The launch of twin \textit{STEREO} spacecraft in 2006 and their subsequent observations of CMEs from the Sun to the Earth have revolutionized the understanding of propagation of CMEs. However, since the discovery and detection of CME in 1971 from a coronagraph onboard OSO-7 (\cc{Tousey 1973}), thousands of CMEs have been observed from a series of space-based coronagraphs e.g., Apollo Telescope Mount onboard Skylab (\cc{Gosling \etal 1974}), Solwind coronagraph onboard \textit{P78-1} satellite (\cc{Sheeley \etal 1980}), Coronagraph/Polarimeter onboard \textit{Solar Maximum Mission (SMM)} (\cc{MacQueen \etal 1980}),  LASCO onboard \textit{SOHO} (\cc{Brueckner \etal 1995}). These observations were complemented by white light data from the ground-based Mauna Loa Solar Observatory (MLSO) K-coronameter which had a FOV from 1.2 \textit{R$_\odot$}-2.9 \textit{R$_\odot$} (\cc{Fisher \etal 1981}) and emission line observations from the coronagraphs at Sacramento Peak, New Mexico (\cc{Demastus \etal 1973}) and Norikura, Japan (\cc{Hirayama \& Nakagomi 1974}).

Although the CMEs were formally discovered in 1971 (\cc{Tousey 1973}), from a survey of literature it is evident that consequences due to CMEs were noticed well before their discovery. For example, CMEs were observed at larger distances from the Sun in radio via interplanetary scintillation (IPS) observations from the 1960s. However, only around the 1980s, the association between IPS and CMEs could be established (\cc{Hewish \etal 1964}; \cc{Houminer \& Hewish 1972}; \cc{Tappin \etal 1983}). Attempts to observe the CMEs in regions in the inner heliosphere from 0.3 AU to 1.0 AU have also been made from the zodiacal light photometers on the twin Helios spacecraft during 1975 to 1983 (\cc{Richter \etal 1982}). However, this attempt of observing the inner heliosphere was with the extremely limited FOV of zodiacal light photometers. Also, heliospheric imagers as \textit{Solar Mass Ejection Imager} (SMEI) (\cc{Eyles \etal 2003}) onboard the \textit{Coriolis} spacecraft launched early in 2003, have observed several CMEs far from the Sun in the heliosphere.

The observations of CMEs in white-light images, kilometric radio observations from space, and metric radio interplanetary scintillation (IPS) observations from the ground have resulted in several studies. In addition to this,  \textit{in situ} observations of CMEs have also been made for decades (\cc{Klein \& Burlaga 1982}; \cc{Zurbuchen \& Richardson 2006}; \cc{Richardson \& Cane 2010}). The interplanetary scintillation (IPS) techniques are based on the measurements of the fluctuating intensity level of several distant meter-wavelength radio sources. The observations of CMEs using IPS and the estimation of their parameters from several techniques have been described in earlier works (\cc{Hewish \etal 1964}; \cc{Watanabe \& Kakinuma 1984}; \cc{Manoharan \& Ananthakrishnan 1990}; \cc{Bisi \etal 2008}). In the present review, we will focus on the observations of CMEs in white light imaging and \textit{in situ} observations only.

Once a CME leaves the inner corona and start moving into the interplanetary space filled with ambient solar wind medium, it takes the name of interplanetary CME (ICME) which undergoes different morphological and kinematic evolution throughout its propagation (\cc{Dryer 1994}; \cc{Zhao \& Webb 2003}). ICMEs have been identified in \textit{in situ} observations and it was found that their plasma and magnetic field parameters are different from that of the ambient solar wind medium. Although it is possible to record a CME near the Sun and to identify the same in the \textit{in situ} observations, a one to one association between remote and \textit{in situ} observations of CMEs is not always easy to establish. There may be several factors responsible for this which will be discussed in the following sections. It is understood that fast CMEs often generate large-scale density waves out into space which finally steepens to form collisionless shock waves (\cc{Gopalswamy \etal 1998a}). This shock wave is similar to the bow shock formed in front of the Earth's magnetosphere. Following the shock, there is a sheath structure which has signatures of significant heating and compression of the ambient solar wind (\cc{Gopalswamy 2004}; \cc{Manchester \etal 2005}). After the shock and the sheath, the ICME structure is found. In the following sections, we describe the evolution of CMEs as observed from remote imaging and \textit{in situ} observations.

\subsection{Observations of Evolution of CMEs}

The main problem in understanding the evolution of CMEs is our limited knowledge about their physical properties. In addition, remote white-light observations (e.g., coronagraphs and heliospheric imagers) allow tracking the propagation of CMEs, but these observations do not provide information on their magnetic field parameters. Associating the near Earth ICME observed \textit{in situ} by the Advanced Composition Explorer (\textit{ACE}) (\cc{Stone \etal 1998}) and \textit{WIND} (\cc{Ogilvie \etal 1995}) spacecraft with Earth-directed front-side halos CMEs seen in LASCO coronagraph images, several attempts have been made in the past (\cc{Richardson \& Cane 2004}; \cc{Jian \etal 2006}). Such studies have proved to be very difficult because of difficulties in determining the 3D speed of Earth-directed CMEs. Another problem is that an \textit{in situ} spacecraft takes measurements along a certain trajectory through the ICME, therefore it does not provide the global characteristics of CME plasma. \textit{SOHO}/LASCO has detected well over 10$^{4}$ CMEs till date and still continues (\cc{\url{http://cdaw.gsfc.nasa.gov/CME_list/}}) (\cc{Yashiro \etal 2004}). SMEI also observed nearly 400 transients during its 8.5 year lifetime, and it was switched off in September 2011. In the following Section~\cc{\ref{CMEobswl}} and \cc{~\ref{insitu}}, we describe the details of different sets of observations of CMEs.

\subsubsection{Remote Sensing Observations of CMEs in White-light} \hspace{0pt}\\ 
\label{CMEobswl}

In white light images, CMEs are seen due to Thomson scattering of photospheric light from the free electrons of coronal and heliospheric plasma. The intensity of Thomson scattered light has an angular dependence which must be considered for measuring the brightness of CMEs (\cc{Billings 1966}; \cc{Vourlidas \& Howard 2006}; \cc{Howard \& Tappin 2009}). They are faint relative to the background corona, but much more transient, therefore a suitable coronal background subtraction is applied to identify them. The advantage of white light observations over radio, IR or UV observations is that Thomson scattering only depends on the observed electron density and is independent of the wavelength and temperature (\cc{Hundhausen 1993}). Thomson scattering is a special case of the general theory of the scattering of electromagnetic waves by charged particles. Since the wavelength of white-light is smaller than the separation between the charge particles in the corona, and the energy of the white-light photons is lower than the rest mass energy of the particles in the corona, therefore the solar photospheric light gets Thomson scattered by electrons in the corona and solar wind.

\begin{figure}[!htb]
\centering
\includegraphics[scale=0.40]{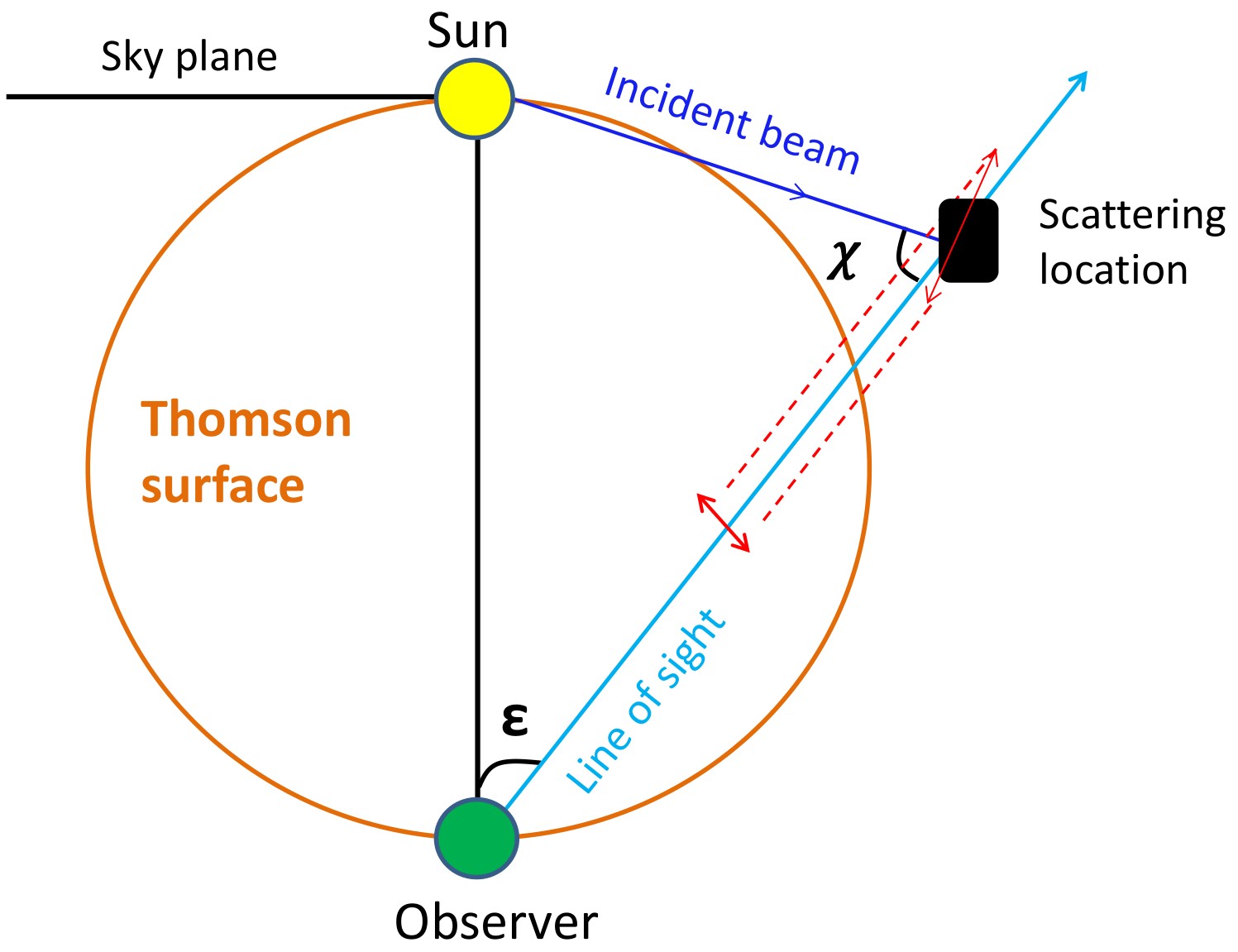}
\caption[Thomson scattering geometry]{The scattering and elongation angles for imaging observations in the context of the Thomson scattering geometry. The line of sight (cyan), incident beam (blue), and the component of scattered intensity seen as parallel to the projected direction of the incident beam (red) are displayed.}
\label{thomson}
\end{figure}

The details of Thomson scattering is given in earlier studies (\cc{Minnaert 1930}; \cc{Billings 1966}; \cc{Howard \& Tappin 2009}; \cc{Howard \& DeForest 2012}; \cc{Howard \etal 2013}). These studies have shown that the received intensity of the scattered light by an observer depends on its location relative to the scattering source and incident beam (Figure~\ref{thomson}). If scattered light is decomposed into two components, then for an observer, the intensity of the component seen as transverse to the incident beam is isotropic, while the intensity of the component seen as a parallel to the projected direction of the incident beam (shown with red in Figure~\ref{thomson}) varies as the square of cosine of scattering angle ($\chi$). The scattering angle is between the vector from scattering location to the observer which is along the line of sight and the vector from scattering source to the center of the Sun which is along the incident beam. It means that $\chi$ is the Sun-scattering location-observer angle. Hence, the efficiency of Thomson scattering measured by an observer is minimum at $\chi$ = 90$\arcdeg$, i.e., on Thomson surface (TS). TS is the surface of a sphere with diameter extending from Sun center to the observer, and all the points of closest approach to the Sun of each line of sight lies on the TS. However, TS is the point where incident light and electron density is found to be maximum. The combined effect of all the three factors is that the TS is the locus of points where the scattering intensity is maximized for a fixed radial distance from the Sun. However, a spread of the observed intensity to larger distances from the TS is noted (\cc{Howard \& DeForest 2012}). This spreading is called `Thomson plateau' which is greater at larger distances (elongations) from the Sun, where elongation ($\epsilon$) is the Sun-observer-scattering location angle. The details of  TS and its theoretical background is discussed in \cc{Howard \& Tappin (2009)}.

It has been shown that the sensitivity of unpolarized heliospheric imagers is not strongly affected by the geometry relative to the TS, and in fact, heliospheric imagers onaboard \textit{STEREO} have observed the CMEs very far from the TS (\cc{Howard \& DeForest 2012}). However, it has also been shown that the polarized brightness measurements of CMEs in the heliosphere, at larger distances from the Sun, are much more localized to the TS than the unpolarized brightness measurements (\cc{Howard \etal 2013}). Conclusively, the observed brightness of a CME can change corresponding to its changing location across the TS and its distance from the Sun, and hence corresponding to observers at different locations. This concept has implications for understanding how the kinematics and morphology of CMEs can appear to be different from observer's perspectives.

\subsubsection{In Situ Observations of CMEs} \hspace{0pt}\\
\label{insitu}

Various plasma, magnetic field and compositional parameters of an ICME are measured by \textit{in situ} spacecraft at the instant when it intersects the ICME. The identification of ICME in \textit{in situ data} is not very straightforward and it is based on several signatures which are summarized below.

\vspace*{5pt}
\paragraph*{\textbf{Magnetic field signatures in the plasma:}} \hspace{0pt}\\

ICMEs are identified in \textit{in situ} observations based on the increased magnetic field strength and reduced variability in the magnetic field (\cc{Klein \& Burlaga 1982}). A subset of ICMEs is known as Magnetic Clouds (MCs) which shows additional signatures such as enhanced magnetic field greater than $\approx$ 10 nT, smooth rotation of magnetic field vector by angles greater than $\approx$ 30$\arcdeg$, and plasma $\beta$ (ratio of thermal and magnetic field energies) less than unity (\cc{Lepping \etal 1990}).

\vspace*{5pt}
\paragraph*{\textbf{Dynamics signatures in the plasma:}} \hspace{0pt}\\

The ICME can be identified \textit{in situ} by its characteristics of expansion during the propagation in the ambient solar wind. Due to expansion, CMEs also show depressed proton temperature in contrast to the ambient solar wind. ICME leading edge, i.e., front has speed greater than its trailing edge and the difference of speeds at boundaries is equal to two times the expansion speed of CME. Hence, a monotonic decrease in the plasma velocity inside an ICME is noticed (\cc{Klein \& Burlaga 1982}). It is also found that the normal solar wind is expected to show an empirical relation between proton temperature and solar wind speed (\cc{Lopez 1987}) as given in Equation~\cc{\ref{eqvswt}}.

\begin{subequations}
\begin{align}
T_{exp} &= (0.031 V_{sw} - 5.1)^{2} \times 10^{3} , \hspace{4pt} when~V_{sw} < 500~km~s^{-1} \\ 
T_{exp} &= (0.51 V_{sw} - 142) \times 10^{3},  \hspace{4pt} when~V_{sw} > 500~km~s^{-1}         
\end{align}
\label{eqvswt}
\end{subequations}

However, it is found that ICMEs do not show the same ``expected'' proton temperature ($T_{exp}$) as it is for the ambient solar wind which can be determined from Equation~\cc{\ref{eqvswt}} . In general, ICMEs typically have proton temperature T$_{p}$ $<$ 0.5 T$_{exp}$ (\cc{Richardson \& Cane 1995}). It is also noted that in an ICME, the electron temperature (T$_{e}$) is greater than proton temperature (T$_{p}$). It is proposed that the ratio of electron to proton temperature, i.e. T$_{e}$/T$_{p}$ $>$ 2 is a good indicator of an ICME (\cc{Richardson \etal 1997}).

\vspace*{5pt} 
\paragraph*{\textbf{Compositional signatures in the plasma:}} \hspace{0pt}\\

The composition of an ICME is different than the ambient solar wind medium. \textit{In situ} observations have shown that alpha to proton ratio (He$^{+2}$/H) is higher ($>$ 6\%) inside an ICME than its values in the normal solar wind. This suggested that an ICME also contains material from the solar atmosphere below the corona (\cc{Hirshberg \etal 1971}; \cc{Zurbuchen \etal 2003}). It is observed that relative to the solar wind, an ICME shows an enhancement in value of $^{3}$He$^{+2}$/$^{4}$He$^{+2}$, heavy-ion abundances (especially iron) and its enhanced charge states (\cc{Lepri \etal 2001}; \cc{Lepri \& Zurbuchen 2004}). ICME associated plasma with enhanced charge states of iron suggests that CME source is ``hot'' relative to the ambient solar wind. It is also noted that ICME shows relative enhancement of O$^{+7}$/O$^{+6}$ (\cc{Richardson \& Cane 2004}; \cc{Rodriguez \etal 2004}). However, few CMEs have been identified with unusual low ion charge states such as the presence of singly-charged helium abundances well above solar wind values (\cc{Schwenn \etal 1980}; \cc{Burlaga \etal 1998}; \cc{Skoug \etal 1999}). Such low charge states suggest that the plasma may be associated with the cool and dense prominence material (\cc{Gopalswamy \etal 1998b}; \cc{Lepri \& Zurbuchen 2010}; \cc{Sharma \& Srivastava 2012}).

\vspace*{5pt} 
\paragraph*{\textbf{Energetic particles signatures in the plasma:}} \hspace{0pt}\\

ICMEs have loops structures rooted at the Sun, therefore the presence of bidirectional beams of suprathermal ($\geq$ 100 eV) electrons (BDEs) is considered as a typical ICME signature (\cc{Gosling \etal 1987}). Sometimes such BDEs are absent when the ICME field lines in the legs of the loops reconnect with open interplanetary magnetic field lines. In addition, the short-term (few days duration) depressions in the galactic cosmic ray intensity and the onset of solar energetic particles are well associated with ICMEs (\cc{Zurbuchen \& Richardson 2006}).

\vspace*{5pt} 
\paragraph*{\textbf{Association with shock and sheath:}} \hspace{0pt}\\

It is understood that some of the fast CMEs generate a forward shock ahead of them. Such shocks are wide and span several tens of degrees in heliospheric longitude, approximately two times the value of the angular width of the driver ICME (\cc{Richardson \& Cane 1993}). In \textit{in situ} observations, a forward shock is identified based on a simultaneous increase in the density, temperature, speed and magnetic field in the plasma. The shock is followed by a sheath region before the ICME/MC. These sheaths are identified as turbulent and compressed regions of solar wind having strong fluctuations in magnetic fields which last for several hours (\cc{Zurbuchen \& Richardson 2006}). The magnetic fields in the compressed sheath region may be deflected out of the ecliptic by draping around the ICME (\cc{McComas \etal 1989}). The compressed and deflected magnetic field in the sheath result in geoeffectiveness. If the pre-shock magnetic field vector in the sheath region makes an angle of 90$^\circ$ with the normal to the shock surface, i.e., for perpendicular shock, then the shock lead to stronger compression of the magnetic field in the sheath than that by parallel shocks. The strongly compressed sheath often give rise to more intense geomagnetic storms  (\cc{Jurac \etal 2002}).

Several studies have shown that different ICMEs show different signatures (\cc{Jian \etal 2006}; \cc{Richardson \& Cane 2010}). For example, few ICMEs show signatures of flux ropes while others do not. However, it is still not well understood why a few ICMEs are not observed as flux-ropes in \textit{in situ} data. Similarly, cold filament materiel which is often observed in COR images as a `bright core' following the cavity is rarely observed in \textit{in situ} observations near 1 AU (\cc{Skoug \etal 1999}; \cc{Lepri \& Zurbuchen 2010}).

\begin{figure}[!htb]
\centering
\includegraphics[scale=0.35]{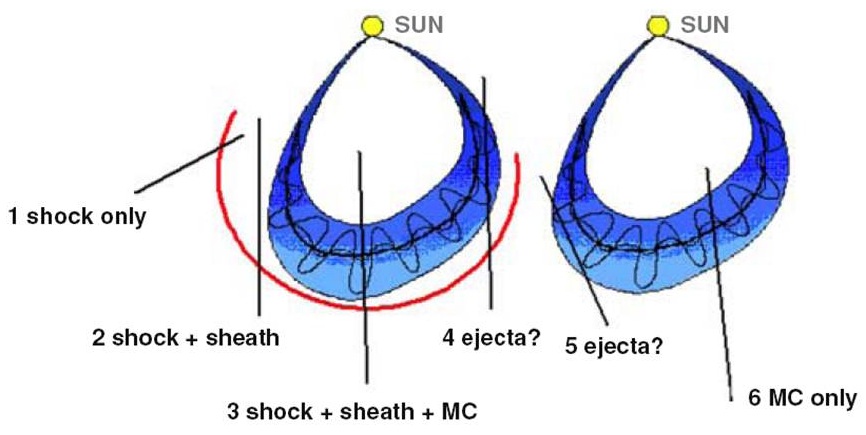}
\caption[Four possible tracks of an \textit{in situ} spacecraft through a CME]{Left panel shows four possible tracks of an observing spacecraft through a CME with a leading shock. Right panel shows another two more tracks through the CME without a leading shock. Track 1 passes through the shock only and track 2 passes through the shock and the sheath of the CME. Track 3 corresponds to a situation when the CME from the Sun is directed exactly towards the \textit{in situ} spacecraft. In this case, the spacecraft measures the shock, sheath, and the magnetic cloud. Tracks 5 and 6 are similar to 4 and 3, respectively, where there is no shock possibly due to a slow speed CME. Trajectories 4 and 5 will not observe the MC structure. (reproduced from \cc{Gopalswamy 2006a})}
\label{insitutraje}
\end{figure}

It is important to mention here that no CMEs show all the signatures and therefore there is no unique scheme to identify them in \textit{in situ} observations. Also, different signatures may appear  for different interval of time and hence, CMEs may have different boundaries in plasma, magnetic field and other signatures. This is possible as different signatures have their origin due to different physical processes. If we identify CMEs based on only a few signatures then they may be falsely identified. Therefore, a practical approach is to identify as many signatures as possible. Such an approach helps for reliable identification of the CMEs in \textit{in situ} observations, however marking of their boundaries may still be ambiguous. \cc{Richardson \& Cane (2010)} have identified approximately 300 CMEs near the Earth during the complete solar cycle 23, i.e.,  between year 1996 to 2009. However, there are some other lists of CMEs observed near the Earth which have compiled slightly differing number of ICMEs based on slightly different criteria (\cc{Richardson \& Cane 1995}; \cc{Cane \& Richardson 2003}; \cc{Richardson \& Cane 2010}).

Before the \textit{STEREO} era, the biggest limitation of CME study was that most of the \textit{in situ} data analysis was restricted to a single point observation at 1 AU while ICMEs are large 3D structures. The limitation of a single point \textit{in situ} observations is illustrated in Figure~\cc{\ref{insitutraje}}. The figure shows how a single point \textit{in situ} instruments can measure different structures and hence show different signatures of an ICME depending on the trajectory of the spacecraft through an ICME.  Such a single point \textit{in situ} spacecraft will also measure the different dynamics of an ICME based on its location within the ICME. Hence, in the absence of information about the part of ICME which is being sampled by the \textit{in situ} spacecraft, it would be difficult to find an association between the speed derived in COR FOV and the one measured \textit{in situ}. Furthermore, since the CMEs evolve during their propagation from the Sun to Earth, making an association between remote observations close to the Sun and \textit{in situ} observations close to the Earth is erroneous. Therefore, multi-point \textit{in situ} observations of an ICME from different viewing perspectives and investigation of the thermodynamic state of CMEs must be carried out.

\subsection{Analysis Methodology for CMEs Kinematics}

Several studies have been carried out to understand the CME kinematics using imaging observations from several space-based instruments (\cc{Schwenn 2006}, and references therein). Among all the space-based instruments dedicated to observing CMEs, the \textit{SOHO}/LASCO launched in 1995 can be considered as the most successful mission in observing thousands of CMEs which led to hundreds of important research papers. \textit{SOHO}/LASCO consists of three nested coronagraphs C1 (no longer operating since June 1998), C2, and C3 that have observed the solar corona from 1.1 \textit{R$_\odot$} to 30 \textit{R$_\odot$}, with overlapping FOVs. Using these observations, several studies were carried out to estimate the source location, mass, kinematics, morphology and arrival times of CMEs (\cc{St. Cyr \etal 2000}; \cc{Xie \etal 2004}; \cc{Schwenn \etal 2005}). Also, to explain the initiation and propagation of CMEs, several theoretical models have been developed  (\cc{Chen 2011}, and references therein). These models differ from one another considerably in the involved mechanism of the progenitor, triggering, and the eruption of a CME. Based on the angular width of a CME observed in coronagraphic images, CMEs were classified as halo, symmetric halo, asymmetric halo, partial halo, limb, and narrow CMEs. Furthermore, based on the acceleration profile of a CME, the CMEs were classified as gradual and impulsive CMEs (\cc{Sheeley \etal 1999}; \cc{Srivastava \etal 1999}). Despite the observations of CMEs with extremely low and high speeds, it is believed that all CMEs belong to a dynamical continuum having no difference in the physics of their initiation process (\cc{Crooker 2002}). With the availability of complementary disk observations of solar active regions and prominences, statistical studies on the association of different types of CMEs with flares and prominences have also been carried out in detail (\cc{Kahler 1992}; \cc{Gopalswamy \etal 2003b}).

\begin{figure}[!htb]
	\centering
		\includegraphics[scale=1.7]{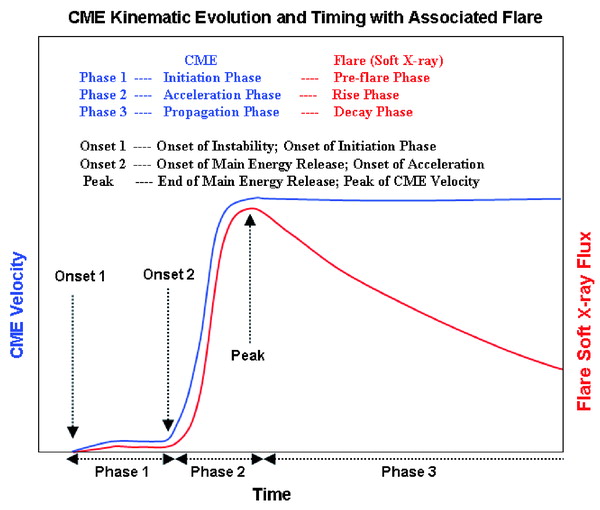}
\caption[Three phase kinematic profile of a CME]{The three different phases of kinematics of a CME and its association with temporal evolution of GOES soft X-ray flux is shown. The initiation, acceleration, and propagation phase of the CME kinematics correspond to the preflare, rise, and decay phase of the associated flare, respectively. (reproduced from \cc{Zhang \& Dere 2006})}
\label{CMEpropimg}
\end{figure}

It is found that a typical CME shows a three-phase kinematic profile: first, a slow rise over tens of minutes, then a rapid acceleration between 1.4 \textit{R$_\odot$}-4.5 \textit{R$_\odot$} during the main phase of a flare, and finally a propagation phase with constant or decreasing speed (\cc{Zhang \& Dere 2006}). These three distinct phases of a CME are shown in Figure~\cc{\ref{CMEpropimg}}. It is noted that after a rapid acceleration phase, the CME accelerates or decelerate slowly in the FOV of coronagraphs (\cc{St. Cyr \etal 2000}; \cc{Yashiro \etal 2004}). The estimated total mass of CMEs range from 10$^{10}$ kg to 10$^{13}$ kg, and the total energy from 10$^{20}$ J to 10$^{26}$ J. The average mass and energy of a CME is 1.4 $\times$ 10$^{12}$ kg and 2.6 $\times$ 10$^{23}$ J, respectively (\cc{Vourlidas \etal 2002b}).

The source locations of the majority of CMEs are within 25$\arcdeg$ from the solar equator, around the solar minimum, although few CMEs are seen at higher latitudes also (\cc{St. Cyr \etal 2000}). Excluding the partial and full halo CMEs, the apparent angular width of CMEs is found to vary from few degrees to more than 120$\arcdeg$, with an average value of about $\approx$ 50$\arcdeg$ (\cc{Yashiro \etal 2004}). These properties derived from a statistical study will also depend on the sensitivity of the coronagraphs and the selection of the sample of CMEs. It is noted that, in the pre-\textit{STEREO} era, the angular width, speed, and mass of CMEs were often estimated from the 2D coronagraphic images of CMEs. Such estimates are subject to the projection and perspective effects. These studies were based on the plane of sky assumption, i.e., CMEs are propagating perpendicular to the Sun-observer line.  Therefore if this assumption of the plane of sky fails, the speed, mass, and energies of CMEs will be underestimated (\cc{Vourlidas \etal 2010}) while the angular width will be severely overestimated (\cc{Burkepile \etal 2004}).

\subsection{Arrival time of CMEs at the Earth}

Realizing the consequences of CMEs on our modern high-tech society, several studies were dedicated at finding a correlation between the intensity of magnetic disturbances on Earth and the characteristics of CMEs estimated near the Sun (\cc{Gosling \etal 1990}; \cc{Srivastava \& Venkatakrishnan 2002}, \cc{2004}). In the context of space weather, understanding the heliospheric evolution of CMEs and predicting their arrival times at the Earth is a major objective of various forecast centers. The prediction of CME/shock arrival time means that forecasters utilize the observables of solar disturbance obtained prior to arrival as inputs to predict whether/when they will arrive. Longer lead time in prediction is yielded if the solar observables are used. The arrival time of CMEs at 1 AU can be related to their characteristics (velocity, acceleration) near the Sun in order to develop the prediction methods for CME’s arrival time. Different kinds of models of CME/shock arrival time prediction have been developed, e.g., empirical models, expansion speed model, drag-based models, physics-based models, and MHD models.

Several studies of evolution of CMEs have been carried out using \textit{SOHO}/LASCO observations, \textit{in situ} observations near the Earth by \textit{ACE} and \textit{WIND} combined with modeling efforts (\cc{Gopalswamy \etal 2000a}, \cc{2001b}, \cc{2005}; \cc{Yashiro \etal 2004}; \cc{Wood \etal 1999}; \cc{Andrews \etal 1999}). These studies were based on the understanding of the kinematics of CMEs using two-point measurements, one near the Sun up to a distance of 30 \textit{R$_\odot$} using coronagraph (LASCO/C2 and C3) images, and the other near the Earth using \textit{in situ} instruments. Using the LASCO images, one could estimate the projected speeds of CMEs, although we lacked information about the 3D speed and direction of the Earth-directed CMEs. These studies, carried out to calculate the kinematics and the travel time of CMEs from the Sun to the Earth, suffered from a lot of assumptions regarding the geometry and evolution of a CME in the interplanetary medium (\cc{Howard \& Tappin 2009}; \cc{Vr\v{s}nak \etal 2010}).

Several models, based on the empirical relationship between measured projected speeds of CMEs and their observed arrival time at 1 AU, have been developed to forecast the CME arrival time at a particular heliocentric distance (\cc{Gopalswamy \etal 2001a}; \cc{Vr\v{s}nak \& Gopalswamy 2002}; \cc{Schwenn \etal 2005}). \cc{Vandas \etal} (\cc{1996}) found that the transit time (in hr) to 1 AU for the CME flux rope (cloud/driver) leading edge is T$_{driver}$ = 85-0.014V$_{i}$ for a slow background solar wind speed (say, 361 km s$^{-1}$), and T$_{driver}$ = 42-0.0041V$_{i}$ for a faster background solar wind speed (say, 794 km s$^{-1}$). Here V$_{i}$ (km s$^{-1}$) is the propagation speed of the leading edge of CME at 18 R$_{\odot}$. Then the transit time of the shock preceding the magnetic cloud is T$_{shock}$ = 74 - 0.015V$_{i}$  for slow solar wind and T$_{shock}$ = 43-0.006V$_{i}$ for fast solar wind. It is found that the difference in time between the CME launch on the Sun and the time when the associated geomagnetic storm reaches its peak is about 80 hr (\cc{Brueckner \etal 1998}).

Among the most typical and widely used prediction models are empirical CME arrival (ECA) and empirical shock arrival (ESA) models. ECA model consider that a CME has an average acceleration up to a distance of 0.7 AU-0.95 AU (\cc{Gopalswamy \etal 2001a}). After the cessation of acceleration, a CME is assumed to move with a constant speed. They found that the average acceleration has a linear relationship with the initial plane-of-sky speed of the CME. The ECA model has been able to predict the arrival time of CMEs within an error of $\pm$ 35 hr with an average error of 11 hr. Later, an empirical shock arrival (ESA) model was able to predict the arrival time of CMEs within an error of approximately $\pm$ 30 hr with an average error of 12 hr (\cc{Gopalswamy \etal 2005}). The ESA model is a modified version of the ECA model in which a CME is considered to be the driver of magnetohydrodynamic (MHD) shocks. The other assumption is that fast mode MHD shocks are similar to gas dynamic shocks. The gas dynamic piston-shock relationship is thus utilized in this model. Various efforts have been made to derive an empirical formula for CME arrival time, based on the projected speed of a large number of CMEs (\cc{Wang \etal 2002}; \cc{Zhang \etal 2003}; \cc{Srivastava \& Venkatakrishnan 2004}; \cc{Manoharan \etal 2004}).

The empirical models adopt relatively simple equations to fit the relations between the arrival time of the CME disturbance at the Earth and their observables such as initial velocity near the Sun. In the majority of these empirical models, the initial speeds of CMEs were measured from plane of sky LASCO/\textit{SOHO} observations and therefore the measured kinematics are not representative of the true CME motion. To overcome plane-of-sky effects, a study of 57 limb CMEs was made to derive an empirical relationship between their radial and expansion speeds as V$_{rad}$ = 0.88V$_{exp}$ (\cc{Dal Lago \etal 2003}). This result led to the use of lateral expansion speed as a proxy for the  radial speed of halo CMEs that could not be measured. Also, in another study of 75 events, an empirical  formula for transit time of CMEs to Earth was derived as, T$_{tr}$ = 203 - 20.77 ln(V$_{exp}$) (\cc{Schwenn \etal 2005}). Their results show that the formula can be used for predicting ICME arrivals, with a 95\% error margin of about 24 hr. Such empirical models have inherent difficulties as they are only math-fit of the measured CME speed and arrival time but do not consider the physics of CME evolution through the ambient solar wind.

Furthermore, a few attempts have been made to fit the observed kinematics profiles of CMEs using an appropriate mathematical function (\cc{Gallagher \etal 2003}). These studies, using \textit{SOHO}/LASCO observations, are subject to large uncertainties due to projection effects. To overcome the projection effects, methods such as forward modeling, which approximates a CME as a cone (\cc{Zhao \etal 2002}; \cc{Xie \etal 2004}; \cc{Xue \etal 2005}) and varies the model parameters to best fit the 2D observations, have been used to estimate the CME kinematics. However, this derived kinematics is also subject to several new sources of errors due to the presumed geometry of the CME. Another method known as polarimetric technique, using the ratio of unpolarised to polarised brightness of the Thomson-scattered K-corona, has been applied to estimate the average line of sight distance of CME from the instrument plane of the sky (\cc{Moran \& Davila 2004}). However, the technique of polarization ratio is only applicable up to $\approx$ 5 \textit{R$_{\odot}$} because beyond this the F-corona cannot be considered as unpolarised.  Thus, the estimation of 3D kinematics of a CME beyond a few solar radii from the Sun was largely undetermined in the pre-\textit{STEREO} era.

Many studies have also shown that CMEs interact significantly with the ambient solar wind as they propagate in the interplanetary medium, resulting in acceleration of slow CMEs and deceleration of fast CMEs toward the ambient solar wind speed (\cc{Lindsay \etal 1999}; \cc{Gopalswamy \etal 2000a}, \cc{2001a}; \cc{Yashiro \etal 2004}; \cc{Manoharan 2006}; \cc{Vr\v{s}nak \& \v{Z}ic 2007}). It was shown that CME transit time depends on both the CME take-off speed and the background solar wind speed. The interaction between the solar wind and the CME is understood in terms of a `drag force' (\cc{Cargill \etal 1996}; \cc{Vr\v{s}nak \& Gopalswamy 2002}). Therefore, the analytical models developed are based on the equation of motion of CMEs where the drag acceleration/deceleration has a quadratic dependence on the relative speed between CME and the background solar wind. It was found that the measured deceleration rates are proportional to the relative speed between CME and the background solar wind, as well as a dimensionless drag coefficient (c$_{d}$) (\cc{Vr\v{s}nak 2001}; \cc{Vr\v{s}nak \& Gopalswamy 2002}; \cc{Cargill 2004}). Recently, a discussion on the variation of the drag coefficient (c$_{d}$) with heliocentric distance was made (\cc{Subramanian \etal 2012}). They adopt a microphysical prescription for viscosity in the turbulent solar wind to obtain an analytical model for the drag coefficient. Furthermore, a simple yet powerful drag-based model (DBM) is developed which can estimate the Sun-Earth transit time of CMEs and their impact speed at 1 AU (\cc{Vr\v{s}nak \etal 2013}). The DBM has also been used widely in the \textit{STEREO} era in several studies as described in Section~\cc{\ref{postst}}

The observations have revealed that the dynamics of CMEs are governed mainly by drag force beyond a certain distance from the Sun. This is perhaps the reason why a few analytical drag-based models (DBM) (\cc{Vr\v{s}nak \& \v{Z}ic 2007}; \cc{Lara \& Borgazzi 2009}; \cc{Vr\v{s}nak \etal 2010}) have been used widely in the literature. However, some earlier studies acknowledge the role of Lorentz force even during the propagation phase of a CME (\cc{Kumar \& Rust 1996}; \cc{Subramanian \& Vourlidas 2005}, \cc{2007}; \cc{Subramanian \etal 2014}). In the direction of modeling efforts, a few numerical MHD simulation models (\cc{Odstrcil \etal 2004}; \cc{Manchester \etal 2004}; \cc{Smith \etal 2009}) have been developed and used to predict CME arrival times (\cc{Dryer \etal 2004}; \cc{Feng \etal 2009}; \cc{Smith \etal 2009}). Despite several studies on CME propagation, using observations combined with models, very little is known about the exact nature of the forces governing the propagation of CME.

A physics-based magnetohydrodynamics (MHD) numerical model is the coupled Wang-Sheeley-Arge (WSA) + ENLIL + Cone model (\cc{Odstrcil \etal 2004}) which has often been used to simulate the propagation and evolution of CMEs in interplanetary space and provides a 1-2 day lead time forecasting for major CMEs (\cc{Taktakishvili \etal 2009}; \cc{Pizzo \etal 2011}). WSA is a quasi-steady global solar wind model that uses synoptic magnetograms as inputs to predict ambient solar wind speed and interplanetary magnetic field polarity at Earth (\cc{Wang \& Sheeley 1995}; \cc{Arge \& Pizzo 2000}). The ENLIL model is a time-dependent, 3D ideal MHD model of the solar wind in the heliosphere (\cc{Odstrcil \etal 2002}, \cc{2004}). The cone model assumes a CME as a cone with constant angular width in the heliosphere (\cc{Zhao \etal 2002}; \cc{Xie \etal 2004}). The input of ENLIL at its inner boundary of 21.5 \textit{R}$_{\odot}$ is taken from the output of WSA to get the background solar wind flows and interplanetary magnetic field.

A physics-based prediction model named ``Shock Time of Arrival'' (STOA) model has been developed based on the theory of blast waves from point explosions. This concept was revised by introducing the piston-driven concept (\cc{Dryer 1974}; \cc{Smart \& Shea 1985}). Another such model is the ``Interplanetary Shock Propagation Model'' (ISPM) which is based on a 2.5D MHD parametric study of numerically simulated shocks. The model demonstrates that the organizing parameter for the shock is the net energy released into the solar wind (\cc{Smith \& Dryer 1990}). The ``Hakamada-Akasofu-Fry version 2'' (HAFv.2) model is a ``modified kinematic'' solar wind model that calculates the solar wind speed, density, magnetic field, and dynamic pressure as a function of time and location (\cc{Dryer \etal 2001}, \cc{2004}; \cc{Fry \etal 2001}, \cc{2007}; \cc{Smith \etal 2009}). This model gives a global description of the propagation of multiple and interacting shocks in nonuniform, stream-stream interacting flows of solar wind in the ecliptic plane. The STOA, ISPM, and HAFv.2 models use similar input solar parameters (i.e., the source location of the associated flare, the start time of the metric Type II radio burst, the proxy piston driving time duration, and the background solar wind speed).

We note that some of the aforementioned models are complicated while others are rather simple and easy, however, no significant differences are found between their prediction capabilities of CME arrival time. The predictions yield a root-mean-square error of $\approx$ 12 hr and a mean absolute error of $\approx$ 10 hr, for a large number of CMEs. Many factors are responsible for the limited accuracies of these models, e.g., (1) The inputs parameters (kinematics and morphology) of the model have their own uncertainties. (2) The real-time background solar wind condition into which CME travels is difficult to either observe or simulate from MHD. (3) The change in the kinematics of the CME due to its interaction with other large or small scale solar wind structures. These factors are difficult to be taken into account in a single model. Improvement in the accuracy of these arrival time models requires a better understanding of both the heliospheric evolution of CME and the ambient solar wind medium. Using the observations of CMEs from instruments onboard \textit{STEREO}, the heliospheric evolution can be better understood by imposing some constraints on the models and methods developed based on the observations from \textit{SOHO}/LASCO.

\section{Studies on CME Propagation in \textit{STEREO} Era}
\label{postst}

The twin \textit{STEREO} (\cc{Kaiser \etal 2008}) spacecraft, launched late in 2006, can observe CMEs in the heliosphere using its identical optical, \textit{in situ} particles, fields and radio instruments on each spacecraft. These instruments are in four different measurement packages named as \textit{Sun Earth Connection Coronal and Heliospheric Investigation} (SECCHI) (\cc{Howard \etal 2008}), \textit{In situ Measurements of PArticles and CME Transients} (IMPACT) (\cc{Luhmann \etal 2008}), \textit{PLAsma and SupraThermal Ion Composition} (PLASTIC) (\cc{Galvin \etal 2008})  and S/WAVES. The IMPACT and PLASTIC packages can provide a chance to measure the \textit{in situ} signatures of CMEs at 1 AU from two vantage points. The suite of instruments in SECCHI package consists of two white light coronagraphs (COR1 and COR2), an Extreme Ultra-violet Imager (EUVI) and two white light heliospheric imagers (HI1 and HI2). The SECCHI package have the capability to continuously image a CME from its lift-off in the corona out to 1 AU and beyond.

The twin \textit{STEREO} spacecraft move ahead and behind the Earth in its orbit with their angular separation increasing by 45$\arcdeg$ per year. The \textit{STEREO} mission overcomes a large observational gap between near Sun remote observations and near-Earth \textit{in situ} observations and provides information on the 3D kinematics of CMEs due to multiple viewpoints on the solar corona. Thus, in the \textit{STEREO} era, the three-dimensional 3D aspects of CMEs could be studied for the first time. Such 3D studies on CMEs was not done in  pre-\textit{STEREO} era  when coronagraphic observations were available only from one location along the Sun-Earth line, as discussed in Section~\cc{\ref{prest}}. Such unique observations led to the development of various 3D reconstruction techniques, e.g., tie-pointing method (\cc{Inhester 2006}), forward modeling (\cc{Thernisien \etal 2009}), etc. Also, several other techniques were developed that are derivatives of the tie-pointing technique: the 3D height-time technique (\cc{Mierla \etal 2008}), local correlation tracking and triangulation (\cc{Mierla \etal 2009}), and triangulation of the center of mass (\cc{Boursier \etal 2009}). These methods have been devised to obtain the 3D heliographic coordinates of CME features in the SECCHI/CORs FOV.

The kinematics of CMEs in 3D over a range of heliocentric distances and their heliospheric interaction have been investigated by exploiting \textit{STEREO}/HI observations (\cc{Davis \etal 2009}; \cc{Temmer \etal 2011}; \cc{Harrison \etal 2012}; \cc{Liu \etal 2012}; \cc{Lugaz \etal 2012}; \cc{Mishra \& Srivastava 2013}, \cc{2014}; \cc{Mishra \etal 2015b}, \cc{2016}). In an effot to combine observations and model, \cc{Byrne \etal} (\cc{2010}) applied the elliptical tie-pointing technique on the COR and HI observations and determined the angular width and deflection of a CME of 2008 December 12. They used the derived kinematics as inputs in the ENLIL model (\cc{Odstr\v{c}il \& Pizzo 1999}) to predict the arrival time of a CME at the L1 near the Earth.

It is noted that the 3D kinematics of CMEs may change beyond the CORs FOV either due to drag forces acting on them or due to CME-CME interaction in the heliosphere. Also, a CME may be deflected by another CME and by nearby coronal holes during its propagation in the heliosphere (\cc{Gopalswamy \etal 2009}). To demonstrate the drag forces acting on the CMEs, \cc{Maloney \& Gallagher} (\cc{2010}) estimated 3D kinematics of CMEs in the inner heliosphere exploiting \textit{STEREO} observations and pointed out different forms of drag force for fast and slow CMEs. The aerodynamic drag force acting on different CMEs will be different and its magnitude will change as the CME propagate in the heliosphere. Therefore, the estimation of the CME arrival time using only the 3D speed estimated from the 3D reconstruction method in COR FOV may not be accurate (\cc{Kilpua \etal 2012}).

In the \textit{STEREO} era, in addition to SECCHI imaging suite, each of the \textit{STEREO} carries its IMPACT and PLASTIC suite which can make the \textit{in situ} observations of the ICMEs. The \textit{in situ} observations of ICMEs from \textit{ACE} and \textit{WIND} spacecraft located along the Sun-Earth line, as well as from \textit{STEREO} located off-Sun-Earth line have been made for several cases (\cc{Rodriguez \etal 2011}; \cc{M\"{o}stl \etal 2014}). Exploiting the \textit{in situ} observations of CME by twin \textit{STEREO}, \cc{Kilpua \etal} (\cc{2009}) suggested that high latitude CMEs can be guided by the polar coronal fields and they can be observed as ICME close to the ecliptic plane. In another study, \cc{Kilpua \etal} (\cc{2011}) emphasized that an ICME cannot be explained in terms of simple flux ropes models because they are observed as different \textit{in situ} structures at both the \textit{STEREO} spacecraft even when the separation between the spacecraft were only few degrees in longitude. Despite the advantage of multi-point \textit{in situ} observations, it is still unclear whether all CMEs have flux ropes or in other words, whether all interplanetary CMEs are magnetic clouds. Also, it is not well understood how a remotely observed CME evolves into an ICME observed \textit{in situ} in the solar wind.

Two recently launched space missions, \textit{Parker Solar Probe (PSP)} in August 2018 (\cc{Fox \etal 2016}) and \textit{Solar Orbiter (SO)} in February 2020 (\cc{M\"{u}ller \etal 2020}), are devoted to revolutionizing our understanding of the solar activity, the corona, solar wind, the generation, acceleration, and transport of solar energetic particles (SEPs). Both \textit{PSP} and \textit{SO} carry a comprehensive suite of \textit{in-situ} and remote-sensing instrumentation. These spacecraft intend to reach much closer to the Sun and perform detailed \textit{in-situ} measurements of nascent solar wind. \textit{PSP} having varying elliptical orbits around the Sun in the ecliptic plane will approach to within 10 \textit{R}$_{\odot}$ from the center of the Sun by 2025. \textit{SO} having highly elliptical and inclined orbits around the Sun will approach to within  0.28 AU from the center of the Sun by 2025. \textit{SO} having increasing orbital tilt will reach 18$\arcdeg$ in the nominal mission (first in March 2025), 25$\arcdeg$ at the start of the extended mission (first in January 2027), and 33$\arcdeg$ in the extended mission (first in July 2029). The \textit{Solar Orbiter Heliospheric Imager} (SoloHI) (\cc{Howard \etal 2020}), Metis coronagraph (\cc{Antonucci \etal 2020}) and the \textit{Wide-field Imager for Solar Probe} (WISPR) (\cc{Vourlidas \etal 2016}) onboard \textit{PSP} will gather images of both the quasi-steady flow and transient disturbances in the solar wind over a large FOV. The differing orbits of the two spacecraft provide two potential of sight through the corona and accelerating solar wind which is further aided by \textit{SOHO}/LASCO along the Sun-Earth line and by \textit{STEREO-A}. There have been several studies exploiting the remote observations of CMEs by \textit{PSP} and \textit{SO} (\cc{Hess \etal 2020}; \cc{Rouillard \etal 2020}; \cc{Laker \etal 2021}). Also, many studies have been reported utilizing the \textit{in-situ} observations of solar wind by \textit{PSP} and \textit{SO} (\cc{McComas \etal 2019}; \cc{Horbury \etal 2020}; \cc{Lavraud \etal 2020}). These two missions promise to revolutionize our understanding of the Sun-heliosphere system, but the results from these missions are not included in the present review. Instead, we focus on the heritage of the recent \textit{STEREO} era providing unprecedented imaging observations from multiple viewpoints that have led to the development of several algorithms and software tools in the last 15 years. The following section focus on the importance of deriving 3D morphology, kinematics, and arrival times of large-scale solar wind structures.

\subsection{Remote Observations of CMEs in the Heliosphere}
In the following, we will focus the white-light imaging observations from only CORs and HIs onboard \textit{STEREO}.

\subsubsection{SECCHI/COR observations} \hspace{0pt}\\

As mentioned earlier SECCHI has two white-light coronagraphs, COR1 is a Lyot internally occulting refractive coronagraph (\cc{Lyot 1939}) and its field of view (FOV) is from 1.4 \textit{R}$_{\odot}$ to 4.0 \textit{R}$_{\odot}$. The internal occultation enables better spatial resolution closer to the limb.  COR1 has a resolution of 7.5$\arcsec$ per pixel with a cadence of 8 min. Another coronagraph, COR2 is an externally occulted Lyot coronagraph similar to LASCO-C2 and C3 coronagraphs onboard SOHO spacecraft with FOV from 2.5 \textit{R}$_{\odot}$ to 15 \textit{R}$_{\odot}$. COR2 observes with a cadence of 15 min and with a resolution of 14.7$\arcsec$ per pixel. The brightness sensitivity of COR1 and COR2 is $\approx$ 10$^{-10}$ B$_\odot$ and 10$^{-12}$ B$_\odot$, respectively. The calibration, operation, mechanical and thermal design of COR1 and COR2 coronagraphs are described in (\cc{Howard \etal 2008}).

\begin{figure*}
\begin{center}
\includegraphics[angle=0,scale=.50]{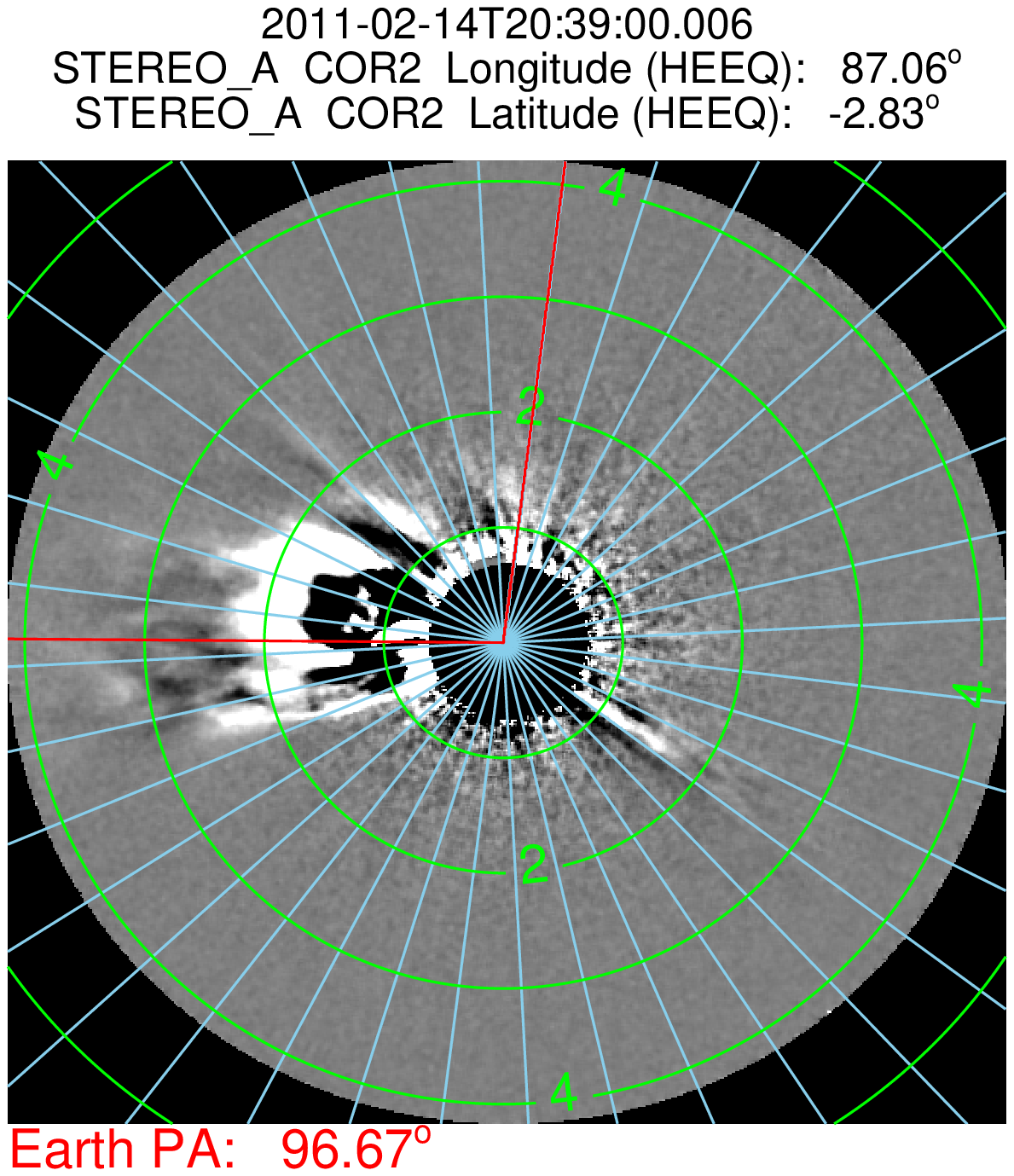}
\includegraphics[angle=0,scale=.50]{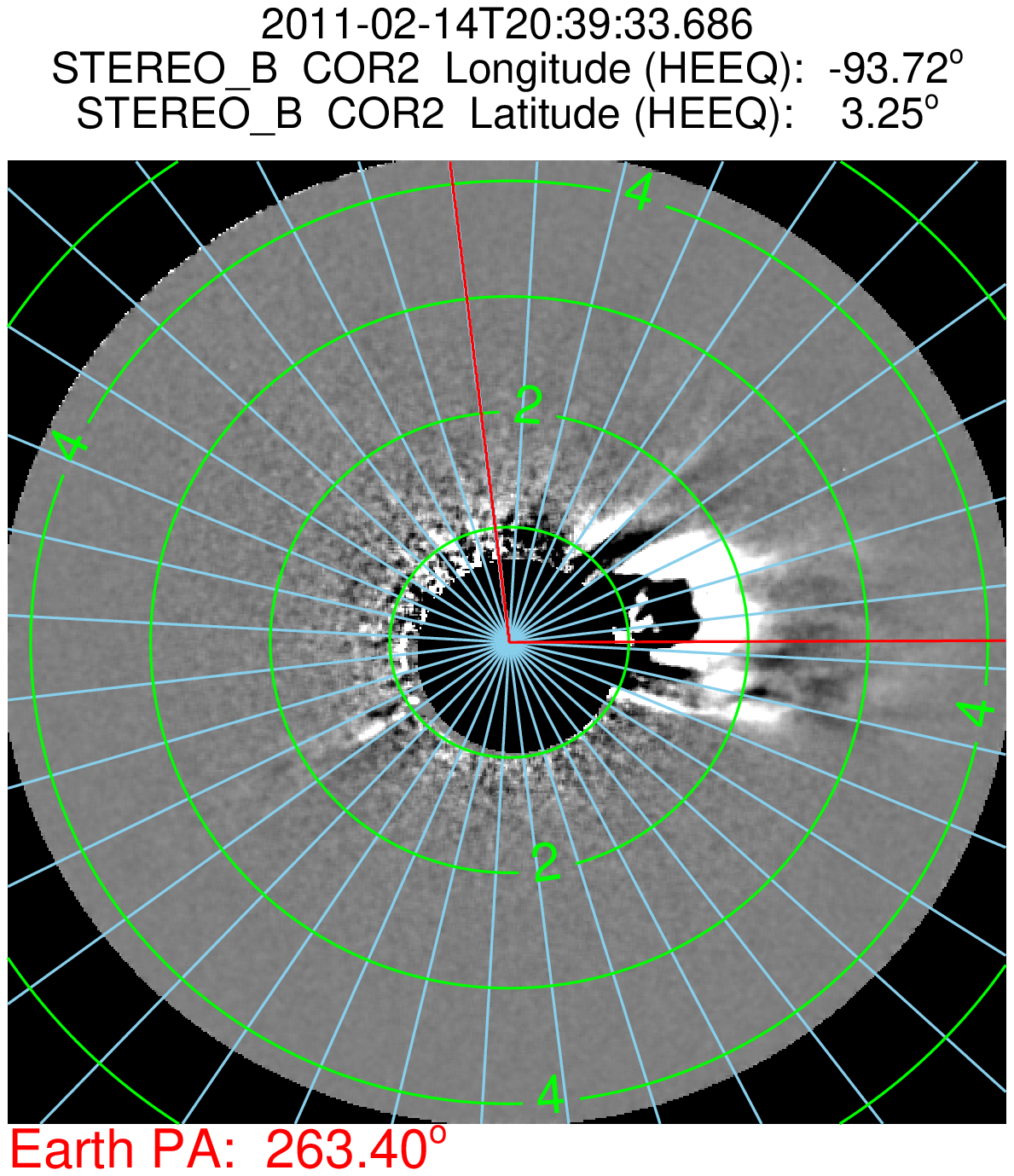}  \\
\includegraphics[angle=0,scale=.50]{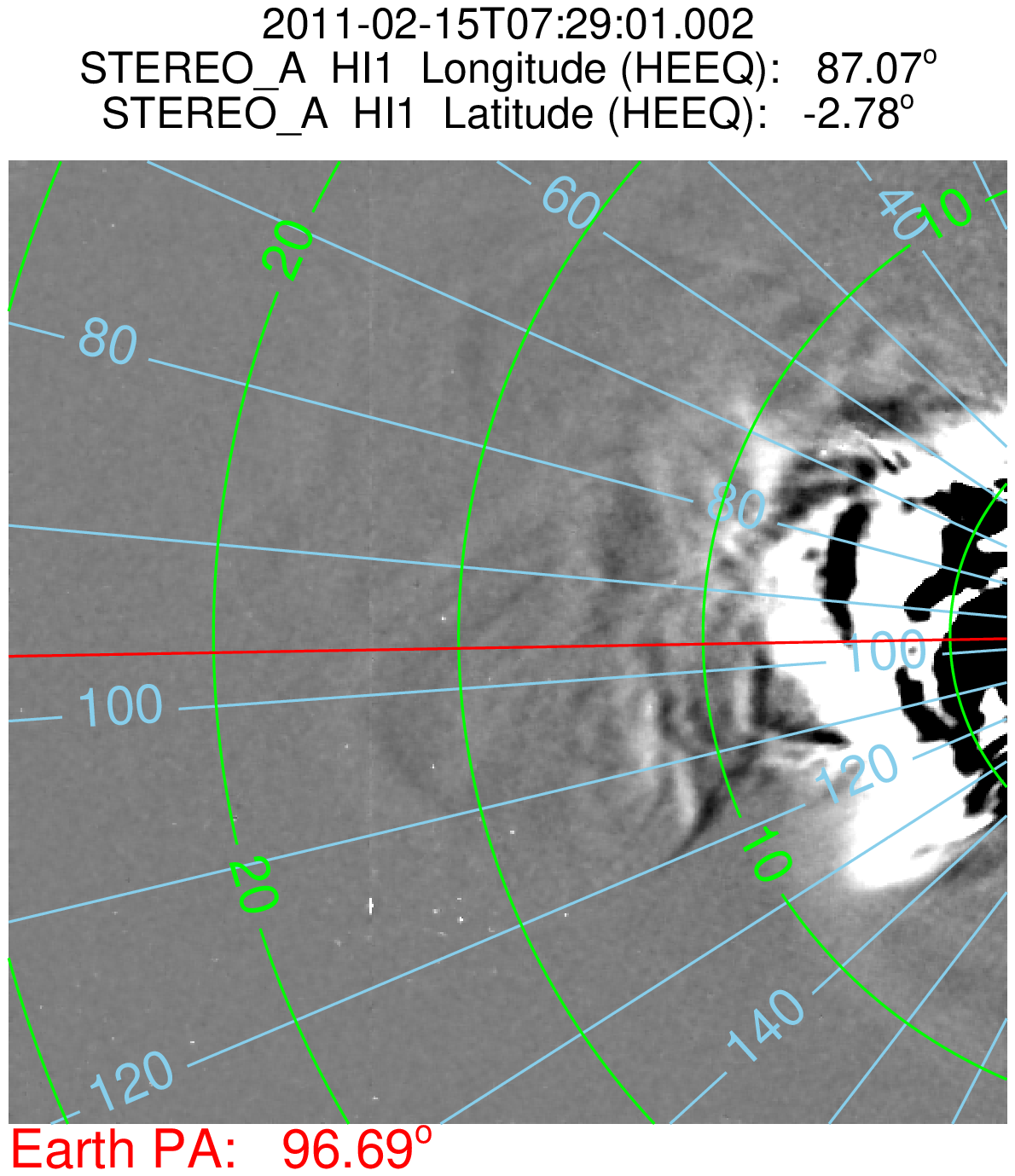}
\includegraphics[angle=0,scale=.50]{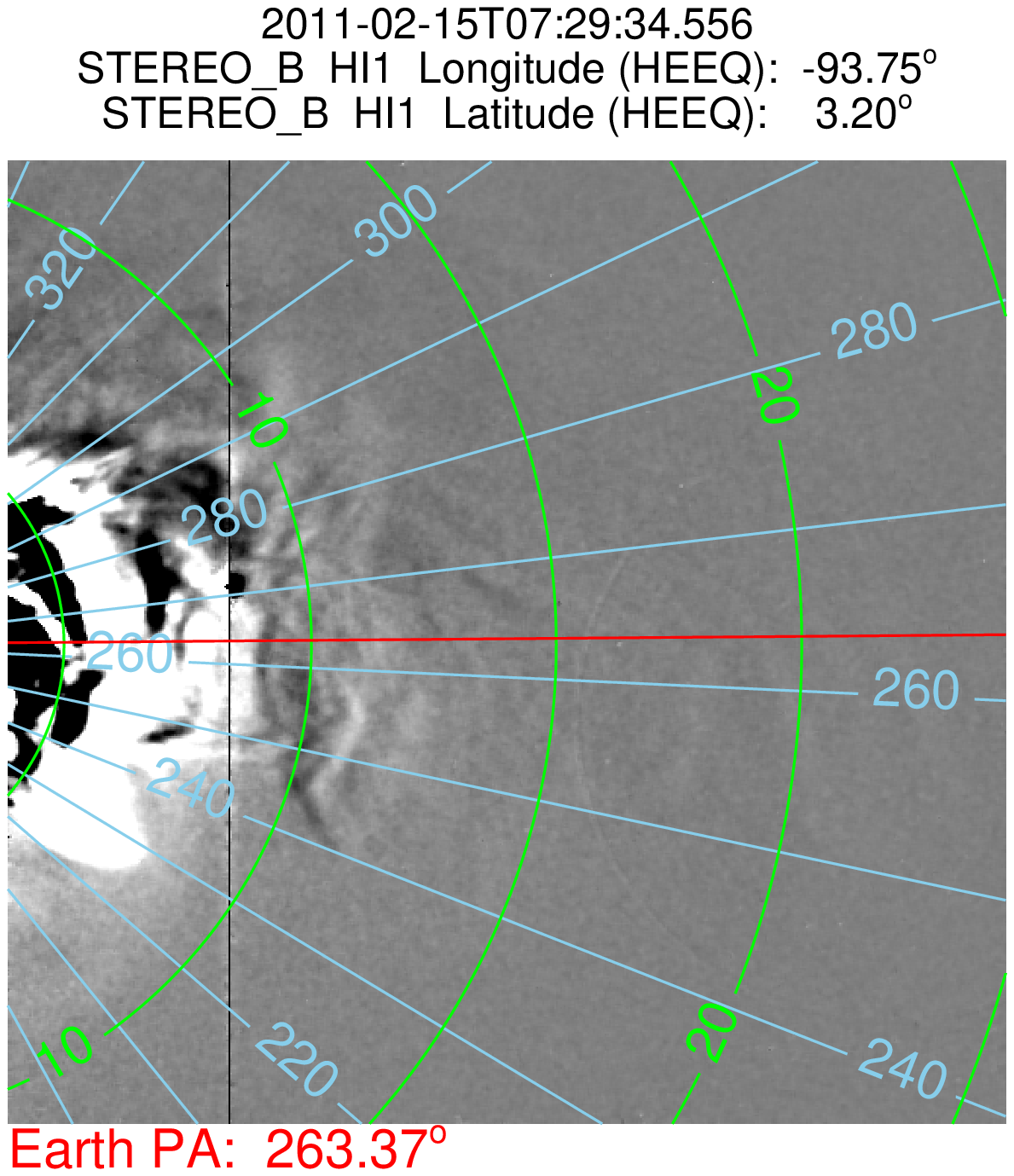}  \\
\includegraphics[angle=0,scale=.50]{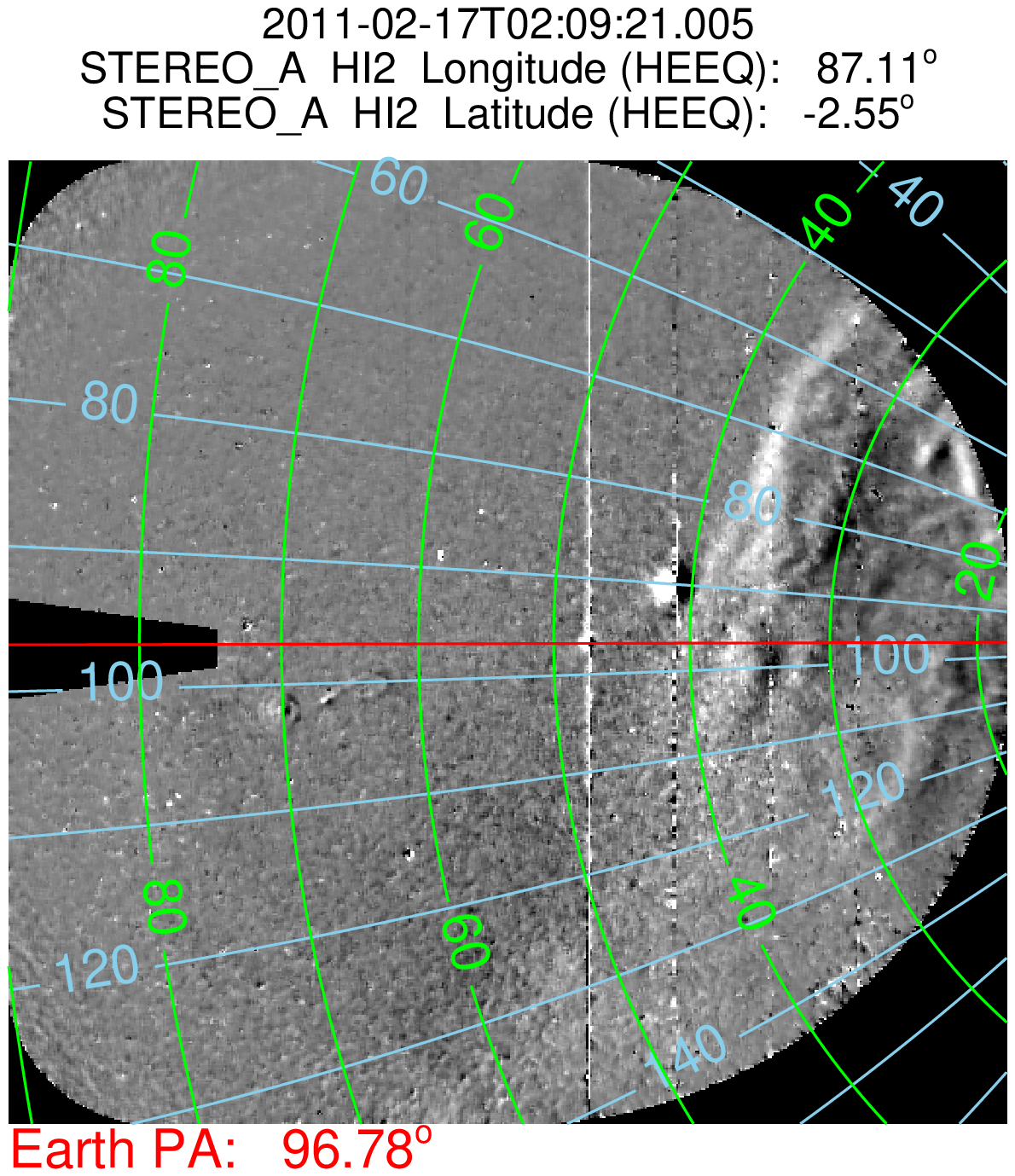}
\includegraphics[angle=0,scale=.50]{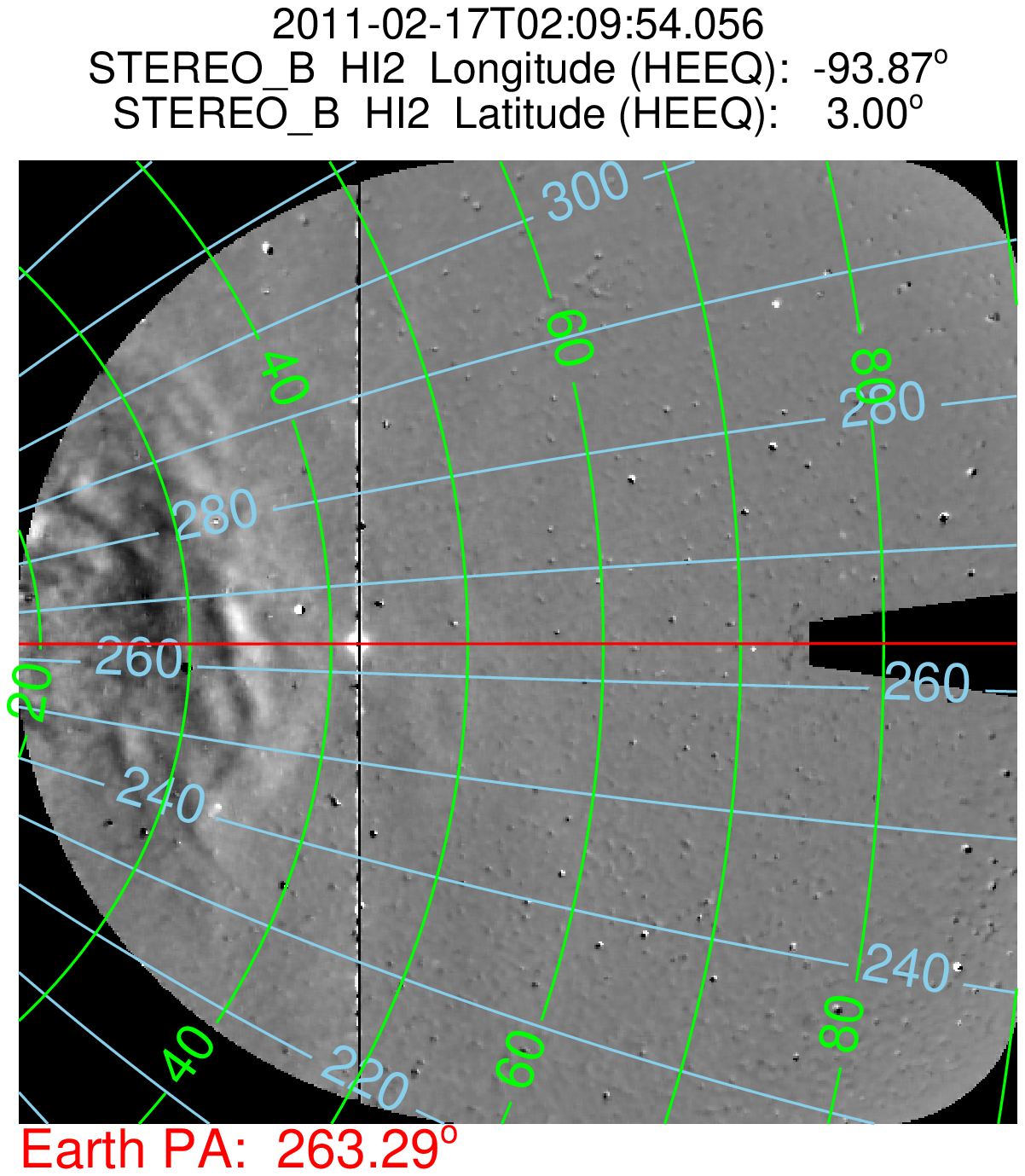}
\caption[Evolution of the CMEs observed during February 2011 in COR2, HI1 and HI2 images]{Evolution of the CMEs observed in COR2, HI1 and HI2 images from \textit{STEREO-A} (left column) and \textit{B} (right column) is shown. The contours  of elongation angle (green) and position angle (blue) are overplotted on the images. The vertical red line in the COR2 images marks the 0$\arcdeg$ position angle contour. The horizontal lines (red) on all panels indicate the position angle of the Earth.}
\label{CMEs_STEREO}
\end{center}
\end{figure*}

\subsubsection{SECCHI/HI observations} \hspace{0pt}\\

SECCHI/Heliospheric Imagers (HIs) detect photospheric light scattered from free electrons in K-corona and interplanetary dust around the Sun (F-corona) similar to CORs. HI also detects the light from the stars and planets within its FOV. The F-corona is stable on a timescale far longer than the nominal image cadence of 40 min and 120 min for the HI1 and HI2 cameras, respectively. The HI1 and HI2 telescopes have an angular FOV of 20$\arcdeg$ and 70$\arcdeg$ and are directed at solar elongation angles of about 14$\arcdeg$ and $\approx$ 54$\arcdeg$ in the ecliptic plane. The HI-A telescopes are pointed at elongation angles to the east of the Sun, whilst HI-B axes are pointed to the west. HI1 and HI2 observe the heliosphere from 4$\arcdeg$-24$\arcdeg$ and 18.7$\arcdeg$-88.7$\arcdeg$ solar  elongation, respectively (\cc{Eyles \etal 2009}). Hence, HI1 and HI2 have an overlap of about 5$\arcdeg$ in their FOVs and therefore permit photometric cross-calibration of the instruments. The HI1 and HI2 are with a resolution of 70$\arcsec$ per pixel and 4$\arcmin$ per pixel, respectively. The brightness sensitivity of HI1 and HI2 is 3 $\times$ 10$^{-15}$ B$_\odot$ and 3 $\times$ 10$^{-16}$ B$_\odot$, respectively (\cc{Eyles \etal 2009}). The images of CMEs observed in the field of view of COR2, HI1, and HI2 are shown in Figure~\cc{\ref{CMEs_STEREO}}. The number of CME ``events'' reported using the HIs onboard \textit{STEREO} is now more than one thousand (\cc{\url{http://www.stereo.rl.ac.uk/HIEventList.html}}), although less than 100 have been discussed so far in the scientific literature (\cc{Harrison \etal 2018}).

\begin{figure*}[!htb]
	\centering
		\includegraphics[scale=0.55]{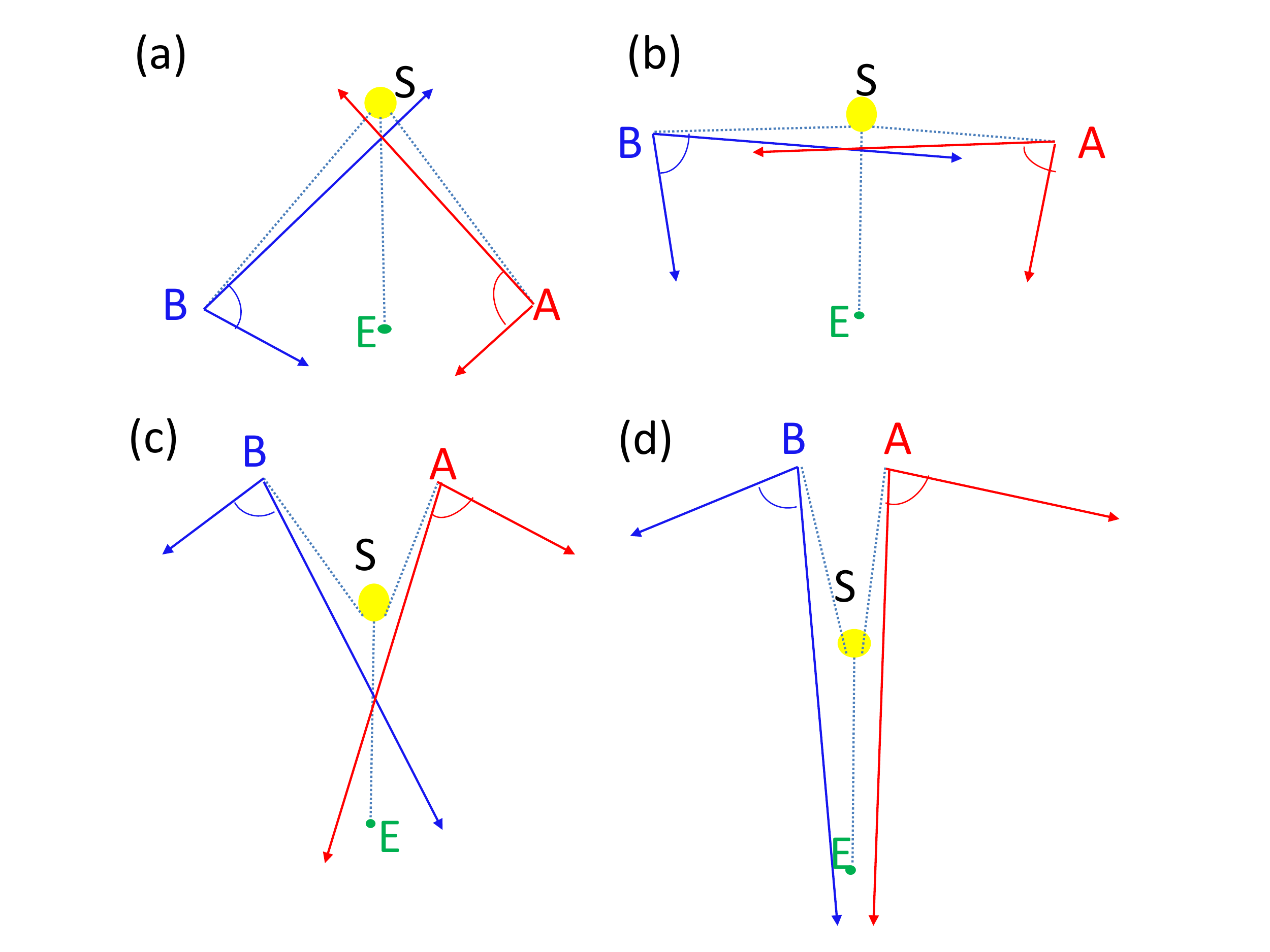}
\caption[Inner and outer edges of HI FOV corresponding to different separation angle of \textit{STEREO-A} and \textit{B}]{Examples of locations of the Sun (yellow), Earth (green), \textit{STEREO-A} (red) and \textit{STEREO-B} (blue) are shown corresponding to different separation angles of twin \textit{STEREO}. The arrows from the \textit{STEREO-A} and \textit{STEREO-B} locations represent the inner (near the Sun) and outer edges of the HI FOV in a qualitative fashion.}
\label{FOVsHIsep}
\end{figure*}

It must be emphasized that HI-A and HI-B view from two widely separated spacecraft at similar planetary angles (Earth-Sun-spacecraft), thus providing a stereographic view. Figure~\cc{\ref{FOVsHIsep}}(a) shows the overall FOVs of HI instruments projected onto the ecliptic plane. The two line of sight drawn with arrows from both \textit{STEREO-A} (red) and \textit{STEREO-B} (blue) spacecraft represent the inner and outer edges of FOVs of HI. The region of the heliosphere observed in the common FOV of HI-A and HI-B only will have a stereoscopic view from \textit{STEREO}. It is also clear from this figure that a CME directed towards the Earth can be observed continuously from the Sun to Earth and beyond from both HI-A and HI-B telescopes. In this scenario, a CME directed eastward from the Earth and \textit{STEREO-B} can only be observed in HI-A FOV but not in HI-B FOV. Similarly, a CME directed westward from the Earth and \textit{STEREO-B} will be observed only in HI-B FOV but not in HI-A FOV.

From Figure~\cc{\ref{FOVsHIsep}}, it can be noted that as the separation (summation of longitude of both \textit{STEREO}) between the \textit{STEREO-A} and \textit{STEREO-B} increases with time, the region of the heliosphere observed simultaneously by both HI-A and HI-B also changes.  From Figure~\cc{\ref{FOVsHIsep}}(b), it is clear that separation between \textit{STEREO-A} and \textit{B} was approximately 175$\arcdeg$ around December 2010, any Earth-directed CMEs during that time cannot be observed near the Sun. They can be observed only a little far from the Sun by both HI-A and HI-B. Therefore, the continuous (Sun to Earth) tracking of CMEs is not possible in this case. Figure~\cc{\ref{FOVsHIsep}}(c) shows that the \textit{STEREO} spacecraft are behind the Sun from the Earth's perspective, i.e., the separation between them is greater than 180$\arcdeg$, HI-A and HI-B will not provide continuous coverage between the Sun and Earth along the ecliptic. Hence, in this scenario also, an Earth-directed CME will not be observed for a significant distance close to the Sun. The other issue of `detectability' of a CME arises when the \textit{STEREO} spacecraft are behind the Sun. In this case, if the CME is directed toward the Earth then it is substantially far-sided for both the \textit{STEREO} spacecraft. Hence, the distance between the CME and \textit{STEREO} increases with time and also as the CME diffuses with time, therefore its detection is difficult but not impossible. Even in such a scenario, some of the Earth-directed CMEs have been detected well in HI FOV (\cc{Liu \etal 2013}). In Figure~\cc{\ref{FOVsHIsep}}(d), the \textit{STEREO} spacecraft are on the other side of the sun with respect to Earth. In this scenario, the Earth does not appear in HI FOV which implies that any CME propagating toward the Earth will not be observed during its journey from the Sun to the Earth. We highlight that the communication with \textit{STEREO-B} got lost around October 2014 and was re-established for a short duration only in August 2016. It has been out of contact since September 2016; therefore, at present, only \textit{STEREO-A} is operating in the absence of \textit{STEREO-B}. Such a loss of \textit{STEREO-B} has limited the operational potential of the overall \textit{STEREO} mission.

\subsection{Analysis and Methodology for CMEs Kinematics using COR2 observations}
\label{anameth}

Various 3D reconstruction methods have been developed which can be used on SECCHI/COR observations, i.e., for a CME feature close to the Sun. These have been reviewed in (\cc{Mierla \etal 2010}). The most widely used 3D reconstruction techniques on the SECCHI/COR observations of CMEs are the tie-pointing method (\cc{Thompson 2009}; \cc{Inhester 2006}) and forward modeling method (\cc{Thernisien \etal 2009}). These methods are often used to estimate the kinematics of CMEs close to the Sun, i.e., before they enter into the HI FOV.

\subsubsection{Tie-point (TP) reconstruction} \hspace{0pt}\\
\label{tiepoint}

The tie-pointing method of stereoscopic reconstruction is based on the concept of epipolar geometry. The position of two \textit{STEREO} spacecraft and the point to be triangulated defines a plane called epipolar plane (\cc{Inhester 2006}). Since every epipolar plane is seen head-on from both \textit{STEREO} spacecraft, it is reduced to a line in the respective image projection. This line is called epipolar line. Epipolar lines in each image can easily be determined from the observer's position and the direction of observer's optical axes. Any object which lies on a certain epipolar line in one image must lie on the same epipolar line in the other image. This straight forward geometrical consequence is known as epipolar constraint.

Due to the epipolar constraint, finding the correspondence of an object in the contemporaneous images from both spacecraft reduces to finding out correspondence along the same epipolar lines in both images. Once the correspondence between the pixels is found, the 3D reconstruction is achieved by calculating the line of sight  rays corresponding to those pixels and on back tracking them in 3D space. Since the rays are constrained to lie in the same epipolar plane, they intersect at a point on tracking backwards. This procedure is called tie-pointing. The point of intersection of both line of sight gives the 3D coordinates of the identified object or feature in both sets of images. Before implementing the method, the processing of SECCHI/COR2 images and the creation of minimum intensity images and then its subtraction from the sequence of processed COR2 images are carried out as described in earlier studies (\cc{Mierla \etal 2008}; \cc{Srivastava \etal 2009}). This method has a graphical user interface (GUI) in the Interactive Data Language (IDL) software and has been widely used in several studies to estimate the 
3D coordinates of a CME's feature (\cc{Mishra \& Srivastava 2013}; \cc{Mishra \etal 2014}).

\subsubsection{Forward modeling method} \hspace{0pt}\\

\begin{figure*}[!htb]
	\centering
		\includegraphics[scale=0.60]{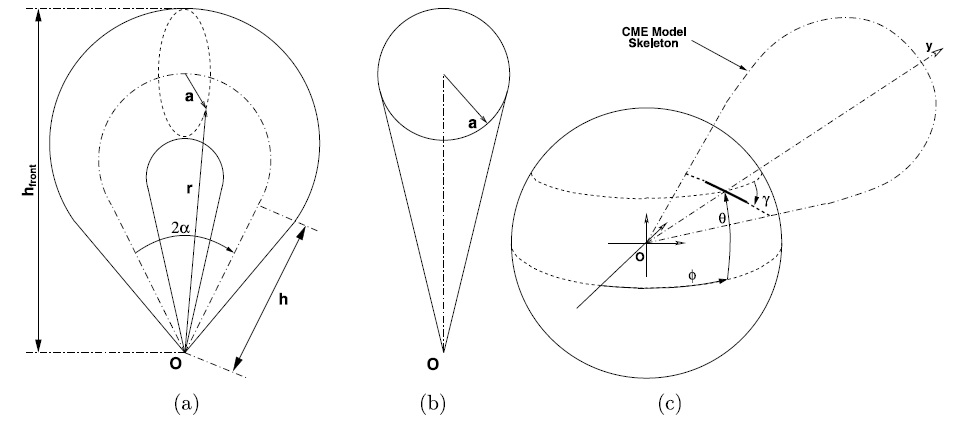}
\caption[GCS model representation]{The left panel (marked as a) shows the Graduated Cylindrical Shell (GCS) model as seen face-on. The central panel (marked as b) shows the GCS model as seen edge-on. The right panel (marked as c) shows the several positioning parameters of the GCS model. The dash-dotted and the solid line represents the axis and a planar cut through the cylindrical shell, respectively. The $\phi$ and $\theta$ are the longitude and latitude of the axis through the centre of the shell, respectively, and $\gamma$ is the tilt angle around the axis of symmetry of the model. (reproduced from \cc{Thernisien \etal 2009})}
\label{modelGCS}
\end{figure*}

In the forward modeling method, a specific parametric shape of CME is assumed and iteratively fit until it matches with its actual image. \cc{Thernisien \etal} (\cc{2009}) developed a method assuming a Graduated Cylindrical Shell (GCS) model to match the CME observed by SECCHI/COR2-A and B. The GCS model represents the flux rope structure of CMEs with two shapes; the conical legs and the curved (tubular) fronts (Figure~\cc{\ref{modelGCS}}). The resulting shape is like a ``hollow croissant''. The model also assumes that the GCS structure moves in a self-similar way. In principle, this technique can also be applied to HI images, however, the technique is widely applied to COR2 images. This is because, in COR2 FOV, the flux-rope structure of CMEs is well identified, while it is not fully developed in COR1 FOV and is too faint in the  HI FOV.

GCS model fitting tool in IDL involves simultaneous adjusting six model parameters so that the resulting GCS flux structure matches well with the observed flux rope structure of the CME (\cc{Thernisien 2011}). These six parameters, including the longitude, latitude, tilt angle of the flux ropes with the height of the legs, half-angle between the legs, and aspect ratio of the curved front are adjusted to match the spatial extent of the CME. These have been discussed in detail in \cc{Thernisien \etal} (\cc{2009}). The best fit six parameters obtained are used to calculate various geometrical dimensions of a CME.

From a space weather perspective, the main advantage of using SECCHI/COR data and the 3D reconstruction methods described above is that it enables estimation of true speed and hence forecasting of the arrival time of CMEs near the Earth with better accuracy. However, information on the deceleration, acceleration or deflection experienced by a CME beyond COR2 FOV cannot be obtained. This may lead to an erroneous arrival time estimation of.

\subsection{Reconstruction methods using COR and HI observations}
\label{Recnsmthd}

It is often observed that when CMEs leave the coronagraphic FOV, the Thomson scattered signal becomes too low to identify a particular feature in both sets of images obtained by \textit{STEREO-A} and \textit{STEREO-B}. Therefore, a method of time- elongation map (\textit{J}-map), initially developed by \cc{Sheeley \etal (1999)} for \textit{SOHO}/LASCO images, is used to track a CME feature in the interplanetary medium. This technique has been implemented on \textit{STEREO}/HI images to reveal the outward motion of plasma blobs in the interplanetary medium (\cc{Rouillard \etal 2009}). In the \textit{STEREO} era, the \textit{J}-maps are now considered necessary for the best exploitation of HI observations to track a CME far away from the Sun (\cc{Davies \etal 2009}; \cc{Harrison \etal 2012}). The details on \textit{J}-maps and its utility to derive the kinematics of CMEs are described in the following Sections~\cc{\ref{Jmapsmthd}}, ~\cc{\ref{SinRcnsMthd}}, ~\cc{\ref{SinRcnsMthdFit}} and ~\cc{\ref{TwinRcnsMthd}}.

\subsubsection{Construction of \textit{J}-maps} \hspace{0pt}\\
\label{Jmapsmthd}

For tracking CMEs in the heliosphere, \textit{J}-maps, also known as time-elongation maps, have often been constructed using long-term background-subtracted running difference images taken from COR2, HI1, and HI2 on \textit{STEREO-A} and \textit{STEREO-B} spacecraft (\cc{Davies \etal 2009}; \cc{Rouillard \etal 2009}; \cc{M\"{o}stl \etal 2011}; \cc{Mishra \etal 2014}). The running difference reveals the changes in electron density between consecutive images. Before computing running differences, the HI image pair is aligned to prevent the stellar contribution in the difference images. This alignment requires precise pointing information of the HI instruments (\cc{Davies \etal 2009}). For this purpose, it is better to use the Level 2 HI data that were corrected for cosmic rays, shutterless readout, saturation effects, flat fields, and instrument offsets from spacecraft pointing. A long-term background image is also subtracted to prepare Level 2 HI data.

\begin{figure}[!htb]
	\centering
		\includegraphics[scale=0.35,angle=90]{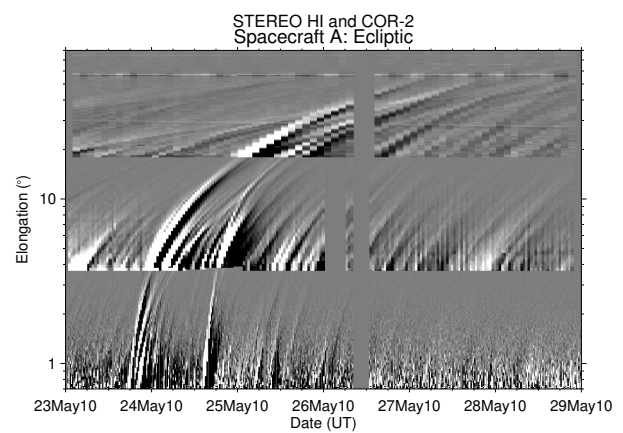}
\caption[\textit{J}-maps using COR2, HI1 and HI2 images]{A typical \textit{J}-map along the ecliptic plane  using running difference images of COR2, HI1 and HI2 on \textit{STEREO-A} spacecraft is shown. The Y-axis shows the elongation angles plotted in logarithmic scale while the X-axis shows the time in UT. Two bright tracks starting on 2010 May 23 at 19:00 UT and May 24 at 14:30 UT represent features of two CMEs and can be tracked up to 50$\arcdeg$ elongation angles.}
\label{JmapMay}
\end{figure}

To construct \textit{J}-maps, \cc{Mishra \etal (2014)} first calculated the elongation and position angles for each pixel of the difference images from COR and HI and extracted a strip of constant position angle along the position angle of the Earth. They considered the position angle tolerance for the COR2 images as 5$\arcdeg$ and 2.5$\arcdeg$ for both HI1 and HI2. Thereafter, they binned the pixels of the extracted strip over a specific elongation angle bin size, viz., 0.01$\arcdeg$ for COR2 and 0.075$\arcdeg$ for both HI1 and HI2. They also took the resistant mean of all pixels over a position angle tolerance in each bin to represent the intensity at a corresponding elongation angle. The resistant mean stacked as a function of time and elongation will produce a time-elongation map (\textit{J}-map). Following this procedure, a typical \textit{J}-map  is shown in Figure~\cc{\ref{JmapMay}} in which the bright curves with positive inclination reveal the propagation of a CME feature.

Using \textit{J}-map, one can track CME features in the heliosphere and derive the elongation-time profile. There have been the development of a plethora of 3D reconstruction techniques which use the time-elongtion profile of a CME to estimate its heliospheric kinematics. These reconstuction techniques are based on different assumtions which make them independent of each other to some degree, as described below. These reconstruction techniques have been applied to a series of images from HIs and have provided information on the evolution of CMEs in the heliosphere (\cc{Howard \etal 2007}; \cc{Davies \etal 2012}; \cc{Mishra \etal 2014}).

\subsubsection{Single spacecraft reconstruction methods} \hspace{0pt}\\
\label{SinRcnsMthd}

It is important to emphasize that when CMEs are very far from the Sun, the `linearity' condition which is imposed on the CMEs near the Sun breaks down. In other words, the `linear assumptions' imposed on an observed CME feature in coronagraphic FOV to convert its measured elongation into the distance are no longer valid when the CMEs are far from the Sun. Near the Sun, the plane of sky assumption is used, i.e., the distance of a feature, d =tan$\alpha$, and further for small $\alpha$; d = $\alpha$ can be used. Such assumptions are not valid when the CME is at a large distance from the Sun. However, if the images of the CMEs are taken at large distances from the Sun and across a large FOV  then, with proper treatment of Thomson scattering and simplistic assumptions about the geometry and trajectory of CMEs, some 3D parameters of CMEs can be estimated by exploiting the images from a single viewpoint alone. Such single spacecraft reconstruction techniques cannot be applied to images obtained from coronagraphs (CORs) as they observe across a small angular extent and therefore the geometric effects of the CME structure are not detectable. As \textit{STEREO}/HI have large FOV and can observe the CMEs at greater distances from the Sun, several attempts have been made to estimate the 3D kinematics of CMEs using single viewpoint observations from HIs. Such single spacecraft reconstruction methods are described below. 

\vspace*{5pt}
\paragraph{\textbf{Point-P (PP) method:}}  \hspace{0pt}\\
\label{PP} 

The Point-P (PP) method was developed by (\cc{Howard \etal 2006}) to convert the elongation angle to distance from the Sun center. This method was developed soon after the launch of SMEI (\cc{Eyles \etal 2003}), and can measure the elongation angle of a moving feature of a CME. The accuracy of this conversion is constrained by the effects of the Thomson scattering process and the geometry of CMEs, which govern their projection in the images. In this method, to remove the plane of sky approximation especially for HIs, it is assumed that a CME is a wide circular structure centered on the sun and an observer looks and tracks the point where the CME intersects the Thomson surface (\cc{Vourlidas \& Howard 2006}). Under these assumptions derived radial distance (R$_{PP}$) of CME from the Sun center is,  R$_{PP}$ = $d_{0}$ $\sin\epsilon$, where $\epsilon$ is the measured elongation of a moving feature and $d_{0}$ is the distance of the observer from the Sun. This method has been used in several earlier studies (\cc{Howard \etal 2007}; \cc{Wood \etal 2009}, \cc{2010}; \cc{Mishra \etal 2014}). In the case where small (elongation) angle approximation can be applied, the PP method is close to the plane of sky approximation.

However, the concept of Thomson surface has been de-emphasized by showing that the maximum intensity of scattered light per unit density is spread over a broad range of scattering angles which is called Thomson plateau (\cc{Howard \& DeForest 2012}; \cc{Howard \etal 2013}). They concluded that CME features can be observed far from the Thomson surface and that their detectability is governed by the location of the feature relative to the plateau rather than the Thomson surface. The existence of this Thomson plateau and the oversimplified CME geometry assumed in the PP method are likely to lead to significant errors in the estimated kinematics of CMEs.

\vspace*{5pt}
\paragraph{\textbf{Fixed-phi (FP) method:}} \hspace{0pt}\\
\label{FP}

\begin{figure*}[!htb]
	\centering
		\includegraphics[scale=8.0]{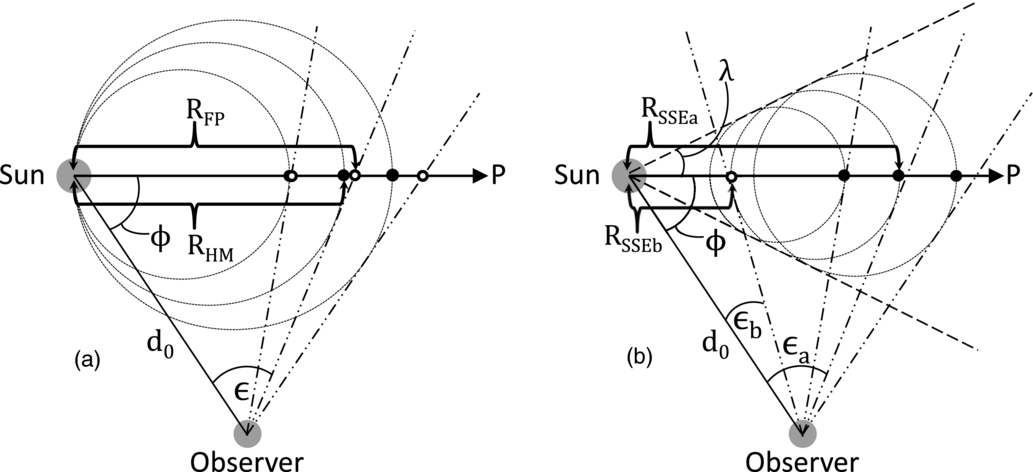}
\caption[CME feature tracking in FP, HM and SSE model geometry]{The left panel (marked as a) shows a tracked CME features in FP (open black dots) and HM (circles/filled black dots) model geometries. The right panel (marked as b) shows the tracked feature corresponding to the geometry of the SSE model. (reproduced from \cc{Davies \etal 2012})}
\label{Davies2012}
\end{figure*}

Analyzing LASCO data, \cc{Sheeley \etal} (\cc{1999}) introduced the concept that the time-elongation map shows an apparent acceleration and deceleration of a CME due to imposed projective geometry on it. However, this effect of apparent acceleration/deceleration was not significant in the LASCO FOV which covers narrow elongation range. After the advent of truly wide-angle imaging with SMEI, \cc{Kahler \& Webb (2007)} developed a  method to convert elongation to radial distance, by assuming that a CME feature can be considered as a point source moving radially outward in a fixed direction ($\phi_{FP}$) relative to an observer located at a distance $d_{0}$ from the Sun (see Figure~\cc{\ref{Davies2012}}a). Using this concept, elongation ($\epsilon(t)$) variation of a moving CME feature can be converted to distance ($R_{FP}(t)$) from the Sun. With these assumptions, the following expression can be derived \cc{(Kahler \& Webb 2007)}. 

\begin{equation}
R_{FP}(t) = \frac{d_{0}\; \sin(\epsilon(t))}{\sin(\epsilon(t)+\phi_{FP})}
\label{FPeqn}
\end{equation}

The fixed radial direction of the propagation of the CME can be determined using the source region of the CME. Also, the initial direction of propagation of a CME can be derived from the 3D reconstruction techniques applicable to COR observations and can be used in Equation~\cc{\ref{FPeqn}}. One major drawback of the FP method is that it does not take into account the finite cross-sectional extent of a CME.

\vspace*{5pt}
\paragraph{\textbf{Harmonic mean (HM) method:}} \hspace{0pt}\\
\label{HM}

To convert elongation angle to radial distance from the center of the Sun, \cc{Lugaz \etal} (\cc{2009}) assumed that a CME can be represented as a self-similarly expanding sphere attached to Sun-center, with its apex travelling in a fixed radial direction. They further assumed that an observer measures the scattered emission from that portion of the sphere where the line of sight intersects tangentially (see Figure~\cc{\ref{Davies2012}}a). Based on these assumptions, they derived the distance (R$_{HM}$) of the apex of the CME from Sun-center as a function of elongation. They found that this distance is the harmonic mean of the distances estimated using the FP and PP methods. Hence, the method is referred to as the HM method.  The distance ($R_{HM}$) of the apex of the sphere from the Sun can be estimated by Equation~\cc{\ref{HMeqn}}  \cc{(Lugaz \etal 2009)}. In the equation, $\phi_{HM}$ is the radial direction of propagation of CME from the Sun-observer line and $\epsilon$ is elongation angle and $d$ is the distance of the observer from the Sun.

\begin{equation}
R_{HM}(t) = \frac{2d_{0}\; \sin(\epsilon(t))}{1 + \sin(\epsilon(t)+\phi_{HM})}
\label{HMeqn}
\end{equation}

Although the spherical geometry of CMEs is included in the Harmonic mean (HM) method, the assumption of such geometry may not be valid at much larger distances from the Sun because of the possible flattening of the CME front during its interaction with the ambient solar wind. The method has been used by \cc{Mishra \etal}(\cc{2014}) where they show that the HM method (based on a propagation direction retrieved from 3D reconstruction of COR2 data) performs better than PP and FP methods.

\vspace*{5pt}
\paragraph{\textbf{Self-similar expansion (SSE) method:}} \hspace{0pt}\\
\label{SSE}

A Self-Similar Expansion (SSE) method represents the elongation variation as a function of time of a CME viewed from a single vantage point (\cc{Davies \etal 2012}). In this method, a CME considered to have a circular cross section, in the plane corresponding to the position angle (PA) of interest, is not anchored to the Sun and, during its propagation away from the Sun, its radius increases such that it always subtends a fixed angle to the Sun center (see Figure~\cc{\ref{Davies2012}}b). They also showed that the SSE geometry can be characterized by an angular half-width ($\lambda$) and in its extreme forms, the SSE geometry is equivalent to the FP ($\lambda$ = 0$\arcdeg$) and HM methods ($\lambda$ = 90$\arcdeg$). It must be noted that $\lambda$ can also be considered as a parameter related to the curvature of the CME front.The distance ($R_{SSE}$)  of a feature using this method at a certain elongation measured from \textit{STEREO-A} or \textit{STEREO-B} can be calculated from the Equation~\cc{\ref{SSEeqn}} (\cc{Davies \etal 2012}).

\begin{equation}
R_{SSE}(t) = \frac{d_{0}\; \sin(\epsilon(t)) (1+\sin(\lambda))}{\sin(\epsilon(t)+ \phi_{SSE}) +\sin(\lambda)}
\label{SSEeqn}
\end{equation}

In all the single spacecraft methods described above, i.e. FP, HM, and SSE, it is assumed that a CME propagates along a fixed radial trajectory (estimated in COR FOV), ignoring real or ``artificial''(see later) heliospheric deflections. Neglecting deflections will induce errors particularly for slow speed CMEs that are more likely to undergo real deflection in the IP medium (\cc{Wang \etal 2004}; \cc{Gui \etal 2011}). This assumption is likely to introduce errors. As a CME moves away from the Sun, not only the direction of propagation but also the geometry plays a role (\cc{Howard 2011}). Such a geometrical effect comes into picture because distances are estimated taking into account the part of the CME which makes a tangent with the line of sight. Therefore, as the CME moves far from the Sun, the observer from a certain location cannot estimate the kinematics of the same part of the leading edge of a CME in subsequent images. This is because of the geometrical effect which produces a situation similar to the deflection of CME and is called `artificial deflection'. This effect leads to an overestimation of the distance of the CME from the FP method which is more severe when the CME approaches longer elongation angles. Finally, the assumption of a circular front in HM and SSE methods may not be valid due to possible flattening of the CME front resulting from its interaction with the structured coronal magnetic field and solar wind ahead of the CME.

\subsubsection{Single spacecraft fitting methods} \hspace{0pt}\\
\label{SinRcnsMthdFit}

\paragraph{\textbf{Fixed-phi fitting (FPF) method:}} \hspace{0pt}\\
\label{FPF}

The original concept of \cc{Sheeley \etal} (\cc{1999}) about deceptive acceleration or deceleration of a CME moving with constant speed in the imager (SMEI $\&$ HI) at large elongation angles from the Sun is used widely to assess the direction of propagation and speed of CME (\cc{Rouillard \etal 2008}; \cc{Sheeley \etal 2008}; \cc{Davis \etal 2009}; \cc{M\"{o}stl \etal 2009}, \cc{2010}; \cc{Howard \& Tappin 2009}; \cc{M\"{o}stl \etal 2011}). Under the assumption that a CME is traveling at a constant speed, the shape of the observed elongation-time profile of CME will be different for observers at different locations. Solving the Equation~\cc{\ref{FPeqn}} for the elongation ($\epsilon(t)$) with assumption of constant velocity (v$_{FP}$) of CME along the fixed radial direction ($\phi_{FP}$), gives us Equation~\cc{\ref{FPFeqn}} \cc{(M\"{o}stl \etal 2009)}.

\begin{equation}
\epsilon(t) = \arctan \Big(\frac{v_{FP}(t)\; \sin(\phi_{FP})}{d_{0} - v_{FP}(t) \; \cos(\phi_{FP})} \Big)
\label{FPFeqn}
\end{equation}

From this Equation, the launch time of a CME from the Sun center, i.e., t$_{0FP}$ can also be calculated and for this $\epsilon$(t$_{0FP}$) = 0 will be satisfied. One should calculate the launch time of a CME in the corona, i.e., at an elongation corresponding to heights in the corona. However, to make the calculation simpler, one can consider the launch time at the Sun's center (\cc{M\"{o}stl \etal 2011}). Theoretical elongation variation obtained from Equation~\cc{\ref{FPFeqn}} can be fitted to match closely with the observed elongation variation for an observed CME by finding the most suitable physically realistic combinations of v$_{FP}$, $\phi_{FP}$ and t$_{0FP}$ values. This approach to find the direction of propagation of a CME and its speed is called the Fixed-Phi-Fitting (FPF) method. This method has been applied to transients like CIRs (\cc{Rouillard \etal 2008}) and also on CMEs (\cc{Davies \etal 2009}; \cc{Rouillard \etal 2009}; \cc{Mishra \etal 2014}).

\vspace*{5pt}
\paragraph{\textbf{Harmonic mean fitting (HMF) method:}} \hspace{0pt}\\
\label{HMF}

Based on HM approximation (\cc{Lugaz \etal 2009}) for CMEs, an expression for the variation of elongation angle with time can be obtained (\cc{Lugaz 2010}). Furthermore, following the fitting version of FP method, i.e., Fixed phi fitting (FPF), another new fitting version of HM method (i.e., Harmonic mean fitting) has been developed (\cc{M\"{o}stl \etal 2011}). In HMF method, the time-variation of elongation  angle ($\epsilon$) for a CME of constant speed (v$_{HM}$) propagating along a fixed radial direction ($\phi_{HM}$) can be written as Equation~\cc{\ref{HMFeqn}} (\cc{M\"{o}stl \etal 2011}).

\begin{equation}
\epsilon(t) = \arccos \Big(\frac{-b + a\; \sqrt{a^{2}+b^{2}-1}}{a^{2}+ b^{2}} \Big) 
\label{HMFeqn}
\end{equation}
 
In this equation, $a$ and $b$ are represented as below. 
	
\begin{equation*}
  a = \frac{2d_{0}} {v_{HM}t} - \cos(\phi_{HM}) \qquad\text{and}\qquad b = \sin(\phi_{HM})
\end{equation*}

It must be noted that in case of a limb CME, its flank will be observed in HI FOV because of the Thomson scattering surface. The flank of a CME is relatively closer to the Sun than its apex. HMF method accounts for this effect and estimates the propagation direction always farther away from the observer compared to the direction derived by FPF method.

\vspace*{5pt}
\paragraph{\textbf{Self-similar expansion fitting (SSEF) method:}} \hspace{0pt}\\
\label{SSEF}

Following the trend of FPF and HMF methods as described above, \cc{Davies \etal} (\cc{2012}) derived a method to convert the measured elongation of an outward moving feature into distance based on selection of an intermediate geometry for the CMEs. In the fitting version of the SSE method, the time-variation of elongation angle of a CME can be expressed in Equation~\cc{\ref{SSEFeqn}} \cc{(Davies \etal 2012)}.   

\begin{equation}
\epsilon(t) = \arccos \Big(\frac{-bc + a\; \sqrt{a^{2}+b^{2}-c^{2}}}{a^{2}+ b^{2}} \Big) 
\label{SSEFeqn}
\end{equation}

In this equation, $a$, $b$ and $c$ are represented as below. 

\begin{equation}
\begin{aligned}
a = \frac{d_{0}(1+c)} {v_{SSE}t} ~-~ \cos(\phi_{SSE}) \quad\text{;}\quad b = \sin(\phi_{SSE}) \quad\text{;}\quad \\ \text{and}\quad c = \pm\sin(\lambda_{SSE})
\end{aligned}
\end{equation}

It must be highlighted that FPF and HMF techniques can be used to estimate only the propagation direction, speed and launch time of the CMEs while SSEF can estimate the additional angular half-width ($\lambda_{SSE}$) of CMEs. Thus, implementation of the SSEF technique requires a four-parameter curve fitting procedure with the assumptions that $\phi_{SSE}$, v$_{SSE}$ and $\lambda_{SSE}$ are constant over the complete duration of the time-elongation profile. The $\lambda_{SSE}$ measures the angular extent of the CME in a plane orthogonal to the observer's FOV. If the SSEF is applied to the front, i.e., the apex of CMEs then the positive form of $c$ is used, while for the trailing edge of the CMEs, its negative form is used. Hence, for CMEs propagating in certain directions, identification of the correct form of the equation to use is very important. It has been pointed out that in the case where SSEF can be applied to time-elongation profiles of features at the front and rear of a CME, then their fitted radial speed would differ while other fitted parameters would be the same (\cc{Davies \etal 2012}).  In the SSEF method, the uncertainties arising from the degrees of freedom associated with the four-parameter fit could also be solved by putting constraints on the other parameters, like $\phi_{SSE}$, $\lambda_{SSE}$, and v$_{SSE}$ to reduce the number of free parameters in the fit. Again, we must emphasize that FPF and HMF methods are the special cases of SSEF method corresponding to $\lambda$ = 0$\arcdeg$ and $\lambda$ = 90$\arcdeg$, respectively.

In a comparison of performance of fitting methods, it was found that there is a large error in the estimated directions when these methods are applied to slow or decelerating CMEs. This is most likely due to a breakdown in their inherent assumptions of constant speed and direction (\cc{Mishra \etal 2014}). They also show that HMF and SSEF methods predict more accurate arrival time and transit speed at L1 than that by FPF method. The main advantage of using FPF, HMF and SSEF methods is that these fitting methods are simple and quick to apply in real-time (\cc{M\"{o}stl \etal 2014}; \cc{Mishra \etal 2014}). In addition, these methods can be used for single spacecraft HI observations, i.e., when any one of \textit{STEREO} spacecraft suffers from a data gap. However, a major disadvantage of these fitting methods is that they assume a constant speed and direction of propagation of the CMEs.

\subsubsection{Multiple spacecraft reconstruction methods} \hspace{0pt}\\
\label{TwinRcnsMthd} 

Reconstruction methods can be greatly improved by using simultaneous observations from two different viewpoints. The \textit{STEREO}spacecraft pair, until the loss of \textit{STEREO-B} in 2014, has provided an ideal platform for such studies as it provided two identical instrument suites at the two different viewpoints. Several twin spacecraft reconstruction methods have been developed to determine the 3D characteristics of CMEs using the time-elongation profiles of the features of a CME from observations from both \textit{STEREO-A} and \textit{STEREO-B} viewpoints. These reconstruction methods utilizing observations of the same CMEs from multiple viewpoints can also be applied on the observations taken from different pairs of wide-angle imagers, e.g., \textit{SOHO}/LASCO and \textit{STEREO}/HI, \textit{SOHO}/LASCO and \textit{SO}/HI, \textit{PSP}/WISPR and \textit{SO}/HI, etc. However, far from the Sun, it is difficult to assume that the same feature of a CME can be observed from different viewpoints or even at different locations in the heliosphere. This increases the complexity of the stereoscopic reconstruction techniques and leads to their inherent limitations. The methods which have been widely used in the literature primarily using observations of heliospheric imagers (HIs) onboard twin \textit{STEREO-A} and \textit{STEREO-B} are described below.

\vspace*{5pt}
\paragraph{\textbf{Geometric triangulation (GT) method:}} \hspace{0pt}\\
\label{GTFP}

\begin{figure}[!htb]
	\centering
		\includegraphics[scale=0.4]{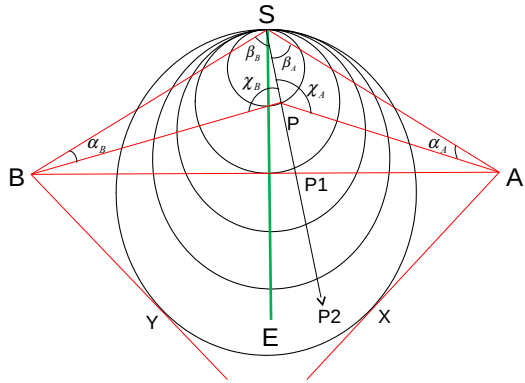}
\caption[Geometric triangulation for a moving CME feature]{Schematic diagram of geometric triangulation for a CME feature moving in the direction of the arrow between the two spacecraft \textit{STEREO-A} and \textit{B}. SE represents the Sun-Earth line and $\alpha$, $\beta$, and $\chi$ denote the elongation, propagation, and scattering angles, respectively. Subscripts A and B represent angles measured from the \textit{STEREO-A} and \textit{STEREO-B} viewpoints. (adapted from \cc{Mishra \etal} \cc{2013})}
\label{GT}
\end{figure}

Based on the concept of triangulation among the two viewpoint of \textit{STEREO} and a CME feature point, a stereoscopic method named as Geometric triangulation (GT) method has been developed (\cc{Liu \etal 2010a}). The GT method assumes that the same feature of a CME can be observed from two different viewpoints and that the difference in measured elongation angles for the tracked feature from \textit{STEREO-A} and \textit{STEREO-B} is entirely due to two viewing directions. Using imaging observations and a Sun-centered coordinate system, the elongation angle of a moving feature can be calculated in consecutive images. The details of the Geometric Triangulation (GT) method in an ecliptic plane applicable for a feature propagating between the two spacecraft have been explained in earlier studies (\cc{Liu \etal 2010a,b}). A schematic diagram for the location of the twin spacecraft and the tracked feature is shown in Figure~\cc{\ref{GT}}. Using this geometry, \cc{Liu \etal} (\cc{2010a}) derived the following Equations:

\begin{equation}
d_{A} =  \frac{r \sin(\alpha_{A} + \beta_{A})}{\sin\alpha_{A}} 
\label{GTeqn1}
\end{equation}
\begin{equation}
d_{B} =  \frac{r \sin(\alpha_{B} + \beta_{B})}{\sin\alpha_{B}} 
\label{GTeqn2}
\end{equation}
\begin{equation}
\beta_{A}+\beta_{B} = \gamma
\label{GTeqn3}
\end{equation}
\\

In the above Equations, $r$ is the radial distance of the feature from the Sun, $\beta_{A}$ and $\beta_{B}$ are the propagation angles of the feature relative to the Sun-spacecraft line. $d_{A}$ and $d_{B}$ are the distances of the spacecraft from the Sun, and $\gamma$ is the longitudinal separation between the two spacecraft. Once the elongation angles ($\alpha_{A}$ and $\alpha_{B}$) are derived from imaging observations, the above equations can be solved for $\beta_{A}$.

\begin{equation}
\beta_{A} = \arctan \Big(\frac{\sin(\alpha_{A}) \sin(\alpha_{B}+\gamma) - f \sin(\alpha_{A}\sin(\alpha_{B})} 
{\sin(\alpha_{A}) \cos(\alpha_{B}+\gamma) + f \cos(\alpha_{A}\sin(\alpha_{B})} \Big)
\label{GTeqn4}
\end{equation} 
\\
where $f$ = $d_{B}$/$d_{A}$ ($f$ varies between 1.04 and 1.13 during a full orbit of the \textit{STEREO} spacecraft around the Sun). Using Equation~\cc{\ref{GTeqn4}}, the propagation direction of a CME can be estimated. Once, the propagation direction has been estimated, the distance of the moving CME feature can be estimated using Equation~\cc{\ref{GTeqn1}}.

In the GT reconstruction method, the effects of Thomson scattering and the geometry of CMEs are not taken into account. However, for Earth-directed events, both view directions
(line-of-sight AP and BP as shown in Figure~\cc{\ref{GT}}) will be nearly symmetrically located from the Sun-Earth line. Therefore, the scattering angles ($\chi_{A}$ and $\chi_{B}$) for both observers will only be slightly different and the resulting difference in the received scattered light intensity for both observers (\textit{STEREO-A} and \textit{STEREO-B}) will be small. The approximation that both observers view the same part of CME may not be true when Earth-directed CMEs are at a large distance from the Sun (for view directions AX and BY as shown in Figure~\cc{\ref{GT}}) and also near the Sun for very wide or rapidly expanding CMEs. It is also rather unlikely that the same feature of a CME will be tracked in each successive image. In light of the aforementioned points, it is clear that the geometry of the CME should be taken into account in any of the reconstruction methods. However, the breakdown of idealistic assumptions about the geometry can result in new errors in the estimated kinematics.

\vspace*{5pt}
\paragraph{\textbf{Tangent to a sphere (TAS) method:}} \hspace{0pt}\\
\label{TAS}

Following the development of the GT method (\cc{Liu \etal 2010a}), another stereoscopic method named as Tangent to a sphere (TAS) (\cc{Lugaz \etal 2010}) was proposed for the reconstruction of CMEs using HIs observations. The TAS method assumes that the CME has a circular cross-section anchored at the Sun and twin \textit{STEREO} observe the tangent to the circular CME front, in contrast to the assumption made in the GT method that the CME is a point. Hence the observers from two viewing locations of \textit{STEREO} do not observe the same CME feature. Under HM approximation, the measured diameter (i.e., $R_{A}$ and $R_{B}$) of the CME from \textit{STEREO-A} and \textit{STEREO-B}, respectively, can be solved for $R_{A}$ = R$_{B}$. The expressions for $R_{A}$ and $R_{B}$ are given in Equations~\cc{\ref{TAS1}} and ~\cc{\ref{TAS2}}, respectively (\cc{Lugaz \etal 2010}).

\begin{equation} 
R_{A}= \frac{2d_{A}\; \sin(\alpha_{A})} {1+\sin(\alpha_{A} + \beta_{A} -\phi_{TAS})} 
\label{TAS1}
\end{equation}

\begin{equation}
R_{B}= \frac{2d_{B}\; \sin(\alpha_{B})} {1+\sin(\alpha_{B}+ \beta_{B} +\phi_{TAS})}
\label{TAS2}
\end{equation}

In the above Equations~\cc{\ref{TAS1}} and ~\cc{\ref{TAS2}}, the parameters $d$, $\alpha$, $\beta$ and $\phi_{TAS}$ are the distance of observer from the Sun, elongation angle, separation angle of observer from the Sun-Earth line, and propagation direction of CME from the Sun-Earth line respectively. The $\phi_{TAS}$ is considered positive in westward direction from Sun-Earth line. The solution of these equations for $\phi_{TAS}$ can be used to estimate the propagation direction of the CMEs. This method to calculate the kinematics of the CME was referred to as tangent-to-a-sphere (TAS) method. This method assumes that measured elongation angle refers to the point where the observers’ line of sight intersects tangentially to the spherical front of the CME.   

\vspace*{5pt}
\paragraph{\textbf{Stereoscopic self-similar expansion (SSSE) method:}} \hspace{0pt}\\
\label{SSSEm}

\begin{figure}[!htb]
	\centering
		\includegraphics[scale=6.5]{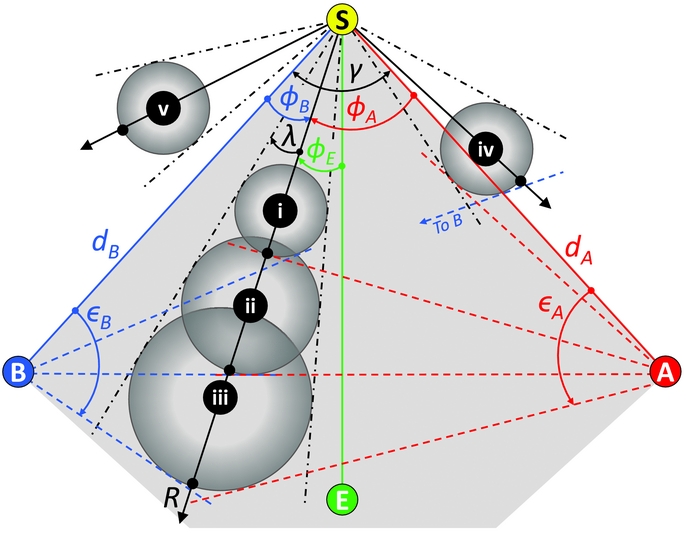}
\caption[The SSE modeled circular CME]{The SSE modeled circular CME, with a constant $\lambda$ is labeled as (i), (ii) and (iii), show three instances of propagation away from the Sun (S) in the common FOV of two observers. The shaded region with gray color represents the common FOV of \textit{STEREO-A} and \textit{STEREO-B}. Geometry marked with (iv) is outside the common FOV however both observers can observe it while geometry (v) is outside the FOV of \textit{STEREO-B} and therefore can only be observed by \textit{STEREO-A}. 
(reproduced from (\cc{Davies \etal 2013})}
\label{SSSE}
\end{figure}

Both GT and TAS methods, as described above, are based on extreme geometrical descriptions of solar wind transients (a point source for GT and an expanding circle attached to the Sun for TAS). Therefore, \cc{Davies \etal} (\cc{2013}) proposed a stereoscopic reconstruction method based on a more generalized SSE geometry, and named it as the Stereoscopic Self-Similar Expansion (SSSE) method. It was shown that the GT and TAS methods can be considered as the limiting cases of the SSSE method. Such a stereoscopic reconstruction with the SSSE method is illustrated in Figure~\cc{\ref{SSSE}}. In this figure, the propagation direction of a CME is shown as $\phi_{A}$ relative to observer \textit{STEREO-A}, $\phi_{B}$ relative to \textit{STEREO-B}, and $\phi_{E}$ relative to Earth (E) and $\gamma$ is the separation angle between both observer located at distances $d_{A}$ and $d_{B}$ from the Sun. At each instance, $\epsilon_{A}$ and $\epsilon_{B}$ is the elongation measured from the line of sight from \textit{STEREO-A} and \textit{STEREO-B}, respectively. The SSSE method is special as we can take a reasonable angular extent ($\lambda$) of CME geometry contrary to the extreme geometrical description taken in both GT and TAS methods. The details of the SSSE method and important considerations for implementation of this method have been discussed earlier (\cc{Davies \etal 2013}; \cc{Mishra \etal 2014}; \cc{Harrison \etal 2017}).

It is also noted that different reconstruction methods, based on different assumptions, often provide different kinematics and arrival time estimates for the CMEs. Therefore, attempts to assess the relative performance of several 3D reconstruction methods, applicable to HI observations, for estimating the arrival time of CMEs, have been made (\cc{Lugaz 2010}; \cc{Howard 2011}; \cc{Mishra \etal 2014}; \cc{Mishra \& Srivastava 2015}). \cc{Mishra \etal} (\cc{2014}) shows that the stereoscopic methods (as described in Section~\cc{\ref{TwinRcnsMthd}}) are more accurate than single spacecraft methods (as described in Section~\cc{\ref{SinRcnsMthd}}) for the prediction of CME arrival times and speeds at L1. Irrespective of the characteristics of the CMEs, among the three stereoscopic methods such as GT, TAS, and SSSE as described before, the TAS method gives the best prediction of transit speed and arrival time within 8 hr for fast CMEs and 17 hr for slow or fast decelerating CMEs. It is also found that the HM method (based on a propagation direction retrieved from 3D reconstruction of COR2 data) performs best among the single spacecraft techniques. Independent of the characteristics of the CMEs, \cc{Mishra \etal} (\cc{2014}) have shown that the HMF and SSEF single spacecraft fitting methods perform better than FPF. All three fitting methods give reasonable arrival time predictions for the fast speed CME that undergoes no discernible deceleration. However, for the slow CME and the fast but decelerating CME, the fitting methods are only accurate within 30 hr in terms of their arrival time prediction and yield relatively larger errors (up to hundreds of km s$^{-1}$) in predicted speed.

\subsection{Drag based model for propagation of CMEs}
\label{DBM}

In both the pre- and post-\textit{STEREO} era, the kinematics of the CME near the Sun has been used either as input to the drag based model or the kinematics is extrapolated to find the arrival time of CMEs at Earth (\cc{Cargill 2004}; \cc{Manoharan 2006}; \cc{Davis \etal 2009}; \cc{Byrne \etal 2010}; \cc{Mishra \& Srivastava 2013}; \cc{Subramanian \etal 2014}). The drag-based model is often used assuming that the Lorentz and gravity forces decrease such that the drag force can largely govern CME dynamics far from the Sun. Although it is not proven that drag is the only force that shapes  CME dynamics in the interplanetary medium, the observed deceleration/acceleration of some CMEs has been closely reproduced by considering only the drag force acting  between the CME and the  ambient  solar  wind medium (\cc{Lindsay \etal 1999}; \cc{Cargill 2004}; \cc{Manoharan 2006}; \cc{Vr\v{s}nak \etal 2009}; \cc{Lara \& Borgazzi 2009}). 
In the \textit{STEREO} era, with the formulation of several 3D reconstruction methods, the 3D kinematics of CMEs estimated in COR2 and HI FOV is used to estimate their arrival time at Earth (\cc{Mishra \& Srivastava 2013}; \cc{Mishra \etal 2014}; \cc{Mishra \etal 2015}). In these studies, the drag based model (DBM) of 
\cc{Vr\v{s}nak \etal}(\cc{2013}) is used to derive the kinematic properties. The DBM is used only for the distance range during which a CME could not be tracked in the \textit{J}-maps constructed from HIs observations.

The DBM model assumes that, after 20 \textit{R}$_\odot$, the dynamics of CMEs is solely governed by the drag force and that the drag acceleration has the form, $a_{d}$ = -$\gamma$ $(v-w)$ $|(v-w)|$, (see, \cc{Cargill \etal 1996}; \cc{Cargill 2004}; \cc{Vr\v{s}nak \etal 2010}), where $v$ is the speed of the CME, $w$ is the ambient solar wind speed and $\gamma$ is the drag parameter. The drag parameter is given by $\gamma$ = $\frac{c_{d}A \rho_{w}}{M + M_{v}}$, where c$_{d}$ is the dimensional drag coefficient, $A$ is the cross-sectional area of the CME perpendicular to its propagation direction (which depends on the CME-cone angular width), $\rho_{w}$ is the ambient solar wind density, $M$  is the CME mass, and $M_{v}$  is the virtual CME mass. The latter is written as, $M_{v}$ = $\rho_{w} V/2$, where $V$ is the CME volume. A statistical study has shown that the drag parameter generally lies between 0.2 $\times$ 10$^{-7}$ and 2.0 $\times$ 10$^{-7}$ km$^{-1}$ (\cc{Vr\v{s}nak \etal 2013}). They assumed that the mass and angular width of CMEs do not vary beyond 20 \textit{R}$_{\odot}$ and also showed that the solar wind speed lies between 300 and 400 km s$^{-1}$ for slow solar wind conditions. For the case where a CME propagates in high speed solar wind or if a coronal hole is present in the vicinity of the CME source region, the ambient solar wind speed should be chosen to lie between 500 and 600 km s$^{-1}$, along with a lower value of the drag parameter.

The DBM can be run instantly which can provide the prediction of ICME expansion and arrival time at any heliospheric locations in the ecliptic plane. It is shown that using a typical value for solar wind speed, the DBM estimate the CME arrival time with typical errors of only around 12 hrs (\cc{Vr\v{s}nak \etal 2013}). We note that DBM ideally assumes that the CME is propagating into an isotropic ambient solar wind. Considering the fact that a CME has actually a 3D structure spanning over different longitudes and latitudes, it is possible that parts of the CME at different latitudes and longitudes are influenced by solar wind of different speeds. One can expect that the high-speed wind from coronal holes may strongly affect those parts of the CME which are at higher latitudes. It may also be the case that a CME experience solar wind of different speeds during the different segments of their heliospheric journey. Such a scenario can arise in the cases when a fast CME encounters a slow CME that was launched earlier in the same direction. It is possible that the performance of DBM and ,thus typical errors in predicting the arrival time of the different portions of the CMEs can be reduced by improving the drawbacks of the simplified drag-based model.

\subsection{Arrival Time of CMEs at the Earth}

\textit{STEREO} observations have greatly enhanced our ability to continuously track CMEs. This is because of \textit{STEREO's} two viewpoints that allow the 3D reconstruction of CMEs. In an attempt to combine the observed CME kinematics with a model, \cc{Kilpua \etal} (\cc{2012}) estimated the 3D speed of CMEs using coronagraphic observations and used it into the CME travel-time prediction models of \cc{Gopalswamy \etal} (\cc{2000a}, \cc{2001a}). They compared the estimated travel time with the actual travel time of CME from the Sun to \textit{STEREO}, \textit{ACE}, and \textit{WIND} spacecraft. They also compared the estimated travel time with that estimated using the projected CME speed into the models. Their study shows that CME 3D speeds give slightly ($\approx$ 4 hr) better predictions than projected CME speeds. However, in their study, a large average error of 11 hr is noted between the predicted and observed travel times.

The large field of view (FOV) of HIs onboard \textit{STEREO} enables the tracking of CMEs to a much larger distance in the heliosphere. Using \textit{STEREO} observations, several attempts have been made to understand the 3D propagation of CMEs and estimate their arrival time (\cc{Mierla \etal 2009}; \cc{Srivastava \etal 2009}; \cc{Kahler \& Webb 2007}; \cc{Liu \etal 2010a}; \cc{M\"{o}st \etal 2011}; \cc{Davies \etal 2012}, \cc{2013}). In a recent study, a CME was tracked beyond the Earth's distance and was shown that a proper treatment of CME geometry must be performed in estimating CME kinematics, especially when a CME is directed away from the observer (\cc{Liu \etal 2013}). Using different reconstruction methods on HI observations, \cc{M\"{o}st \etal} (\cc{2014}) shows an absolute difference between predicted and observed CME arrival times of 8.1 $\pm$ 6.3 hr. These studies have shown that longer tracking of CMEs using HIs observations is necessary for improved understanding of their evolution in the heliosphere.

To understand the heliospheric evolution of CMEs from the Sun to Earth, the kinematics of several CMEs have been estimated by implementing suitable 3D reconstruction methods to remote sensing observations of CMEs (\cc{Mishra \& Srivastava 2013}). These studies suggested that the use of reconstruction methods on HI data combined with DBM gives a better prediction of the CME arrival time than using only 3D speed estimated in COR FOV. Thus, near-Sun 3D speed of CMEs with an assumption that the speed remains constant up to L1, can not accurately predict the arrival time for a majority of CMEs. Sometimes CMEs are observed to erupt in quick succession and, under certain favorable initial conditions, can interact or merge with each other during their propagation in the heliosphere (\cc{Harrison \etal 2012}. Therefore, the interaction of CMEs in the heliosphere is expected to be more frequent near the solar maximum. In the \textit{STEREO} era, one focus of the studies has been to understand the propagation of multiple CMEs following one another from the Sun to Earth and their consequences on hitting the Earth's magnetosphere.

The HI observations have helped to witness several cases of interacting CMEs. Many attempts have been made to understand CME-CME interaction at a range of distances from the Sun using SECCHI/HI observations (\cc{Harrison \etal 2012}; \cc{Temmer \etal 2012}; \cc{M\"{o}stl \etal 2012}; \cc{Lugaz \etal 2012}). It has been shown that during the interaction of CMEs, their kinematics may change. Therefore, such interactions complicate the problem of estimating their arrival time, and any space weather prediction scheme estimating the arrival time of interacting CMEs must take their post-interaction kinematics into account. Therefore, it is important to understand the nature of CME-CME collision by measuring the energy and momentum exchange during the collision/interaction of CMEs.

The actual arrival time of remotely tracked CMEs at Earth can be marked using \textit{in situ} observations near 1 AU. The actual arrival time of some geoeffective CMEs can also be inferred by monitoring the geomagnetic perturbations. These actual arrival times can be compared with the arrival times estimated based on the kinematics obtained from reconstruction methods. However, the identification of a CME in \textit{in situ} observations is not straightforward. The difficulty in the identification further increases when the CMEs arrive as structures formed due to interaction or collision of several CMEs (\cc{Burlaga \etal 2001}). As they interact, they experience a change in their plasma, dynamic and magnetic field parameters. Hence, the collision of CMEs may lead to a new type of solar wind structure which is expected to show different \textit{in situ} signatures than the signatures of isolated CMEs. In addition, such new structures might have a different geomagnetic response as compared to isolated CMEs described in Section~\cc{\ref{insitu}}.

In addition, the interaction or collision of successive CMEs can, in some cases, produce an extended period of southward B$_{z}$ and cause strong geomagnetic storms (\cc{Farrugia \etal 2006}). The geomagnetic responses of interacting CMEs have been explored in several studies described in the following Section~\cc{\ref{cme_int}}. In addition, studies have been devoted at understanding the arrival time, \textit{in situ} identification of interacting CMEs at 1 AU, and various plasma processes during the interaction of CMEs that can change the initial identity and properties of CME plasma. In the \textit{STEREO} era, by exploiting the Sun to Earth remote observations of CMEs from twin viewpoints, one expected to have better success in predicting the speed and direction of a CME near Earth. However, from space weather perspectives, without the knowledge about negative B$_{z}$ component of CME, it would remain difficult to predict the intensity of resulting geomagnetic storms well in advance.

\subsection{CME-CME Interaction}
\label{cme_int}

The possibility of CME-CME interaction has been reported much earlier by analyzing \textit{in situ} observations of CMEs by Pioneer 9 spacecraft (\cc{Intriligator 1976}). The compound streams (interaction of CME-CIR or CME-CME) were first inferred by \cc{Burlaga \etal} (\cc{1987}) using observations from \textit{Helios} and \textit{ISEE-3} spacecraft. They showed that such compound streams formed due to interactions have amplified parameters responsible for producing major geomagnetic storms. Using wide field of view coronagraphic observations from LASCO and long-wavelength radio observations, \cc{Gopalswamy \etal} (\cc{2001c}) provided for the first time evidence for CME-CME interaction. \cc{Burlaga \etal} (\cc{2002}) identified a set of successive halo CMEs directed toward the Earth and found that they appeared as complex ejecta near 1 AU (\cc{Burlaga \etal 2001}). They inferred that these CMEs launched successively, merged en route from the Sun to Earth and formed complex ejecta in which the identity of individual CMEs was lost. Thus, these interactions are of great importance from the space weather point of view.

It has also been shown that CME-CME interactions are important as they can result in an extended period of enhanced southward magnetic field which can cause intense geomagnetic storms (\cc{Farrugia \etal 2006}). Such interactions help to understand the collisions between large scale magnetized plasmoids and hence various plasma processes involved. Also, if a shock from a following CME penetrates a preceding CME, it provides a unique opportunity to study the evolution of the shock strength and structure and its effect on preceding CME plasma parameters (\cc{Lugaz \etal 2005}; \cc{M\"{o}stl \etal 2012}; \cc{Liu \etal 2012}).

Estimating the accurate kinematics and arrival time of CMEs at Earth is crucial for predicting space weather effects. Since CME-CME interactions are responsible for changing the kinematics of interacting CMEs, such interactions need to be examined in detail. Furthermore, as the subset of CMEs are identified as MCs which are flux-rope structures, the reconnection between magnetic flux ropes can be explored by studying cases of CME-CME interactions (\cc{Gopalswamy \etal 2001c}; \cc{Wang \etal 2003}). Such reconnection in CME-CME interaction are known to lead to solar energetic particles (SEPs) events (\cc{Gopalswamy \etal 2002}). \cc{Wang \etal} (\cc{2003}) have shown that a forward shock can cause an intense southward magnetic field of long duration in the preceding MC. Such modifications in the preceding cloud are important for space weather prediction.

It was realized well before the era of wide-angle imaging far from the Sun that CME-CME and CME-shock interactions are important candidates to be studied from physics and space weather prediction point of view. In pre-\textit{STEREO} era, the understanding of involved physical mechanisms in CME-CME or CME-shock interaction was achieved mostly from magnetohydrodynamic (MHD) numerical simulations of the interaction of a shock wave with a magnetic cloud (MC) (\cc{Vandas \etal 1997}; \cc{Vandas \& Odstrcil 2004}; \cc{Xiong \etal 2006}), the interaction of two ejecta (\cc{Gonzalez-Esparza \etal 2004}; \cc{Lugaz \etal 2005}; \cc{Wang \etal 2005}), and the interaction of two MCs (\cc{Xiong \etal 2007}, \cc{2009}). However, only a few attempts could be made to understand the CME-CME interaction using imaging observations near the Sun (\cc{Gopalswamy \etal 2001c}) and \textit{in situ} observations near the Earth (\cc{Burlaga \etal 2001}).

In the \textit{STEREO} era, the twin spacecraft observations enabled to determine the 3D locations of CMEs features in the heliosphere and hence provide direct evidence of CME-CME interaction using images from Heliospheric Imagers. However, immediately after the launch of \textit{STEREO}, during deep extended solar minimum, not many interacting CMEs were observed. As the solar cycle 24 progressed, CME interaction appeared to be a fairly common phenomenon, in particular around solar maximum.

In \textit{STEREO} era, several cases of interacting CMEs in the inner heliosphere have been extensively studied using observations and numerical simulations to understand the physical processes occurring during CME-CME interaction. For example, the interacting CMEs of 2010 August 1 have been studied by several researchers using primarily the \textit{STEREO}/HI (white light imaging), near-Earth \textit{in situ} and, \textit{STEREO}/Waves radio observations (\cc{Harrison \etal 2012}; \cc{Liu \etal 2012}; \cc{M\"{o}stl \etal 2012}; \cc{Temmer \etal 2012}; \cc{Mart\'{i}nez Oliveros \etal 2012}; \cc{Webb \etal 2013}). These studies have shown that CME-CME interaction can lead to change in the properties of CMEs, such as their propagation speed, size, expansion speed, direction of propagation, temperature, internal magnetic field, etc. Therefore, understanding such interactions/collisions of CMEs are important for accurate space weather forecasting. Using \textit{STEREO} imaging observations, several key questions that are not well understood regarding CME interaction have been addressed.

\begin{enumerate}
\item{How do the dynamics of CMEs change after interaction? What is the regime of interaction, \textit{i.e.} elastic, inelastic, or super-elastic? (\cc{Lugaz \etal 2012}; \cc{Shen \etal 2012b}; \cc{Mishra \& Srivastava 2014}; \cc{Mishra \etal 2015a}, \cc{2016}, \cc{2017}).} 
\item{What are the consequences of the interaction of CME-shock structure? How does the overtaking shock change the plasma and magnetic field properties of the preceding magnetic cloud? (\cc{Lugaz \etal 2005}, \cc{2012}; \cc{Liu \etal 2012}).}
\item{What are the favorable conditions for the merging of CMEs and the role of magnetic reconnection in it? (\cc{Gopalswamy \etal 2001c}).} 
\item{What is the possibility for the production of a reverse shock at the CME-CME interaction site? (\cc{Lugaz \etal 2005}).}
\item{Do these interacted structures produce different geomagnetic consequences than individual CMEs, on their arrival to magnetosphere? (\cc{Farrugia \etal 2006}).} 
\item{What are the favorable conditions for the deflection of CMEs and enhanced radio emission during CME-CME interaction? (\cc{Lugaz \etal 2012}; \cc{Mart\'{i}nez Oliveros \etal 2012}).} 
\end{enumerate}

\begin{figure}[!htb]
 \centering
  \includegraphics[scale=0.70]{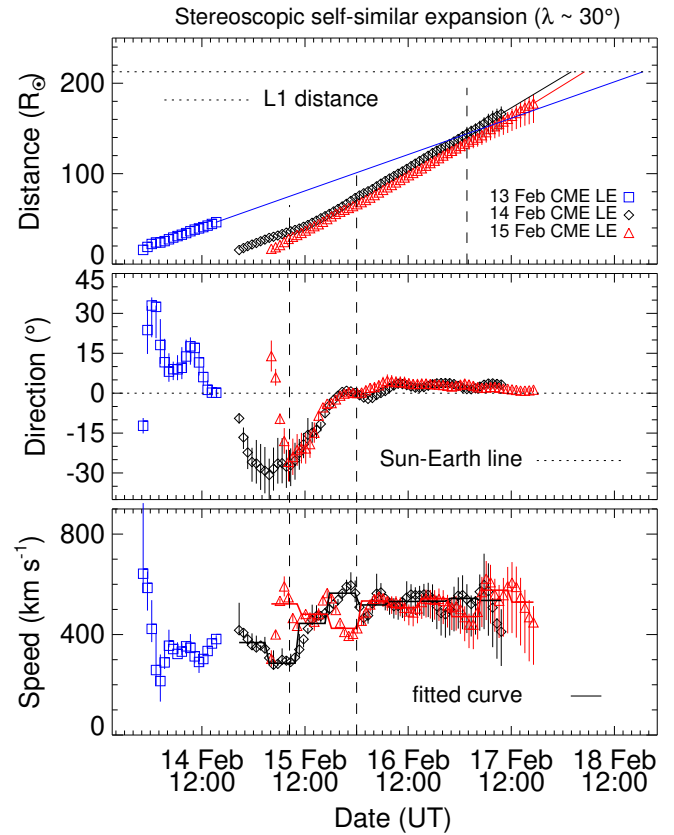}
 \caption[Estimates of distance, propagation direction and speed of the 2011 February 13-15 CMEs using SSSE method]{From top to bottom, the panels show the distance, propagation direction and speed (as obtained using SSSE method) of the leading edge (LE) of CME1 (blue), CME2 (black) and CME3 (red). CME1, CME2, and CME3 were launched on 13, 14 and 15 February 2011 respectively. The horizontal dashed line in the top panel marks the heliocentric distance at the L1 point. The dashed horizontal line in the middle panel marks the Sun-Earth line. The speed shown with symbols is estimated from the differentiation of distance points using three-point Lagrange interpolation. The speed shown with the solid line is determined by differentiating the fitted first order polynomial for estimated distance for each 5 hr interval. From the left, the first and second vertical dashed lines mark the start and end of the collision phase, respectively, for the collision of CME3 and CME2. In the top panel, the rightmost vertical dashed line marks the inferred interaction between CME2 and CME1. The vertical solid lines at each data point show the error bars (adapted from \cc{Mishra \& Srivastava 2014})}.
\label{kin_int_CMEsFeb}
\end{figure}

To understand the interaction of CMEs, a study of \cc{Mishra \& Srivastava}  (\cc{2014}) investigated the signatures of 3 interacting CMEs in remote sensng and \textit{in situ} observations. These three CMEs were observed to have launched from the Sun successively on 13, 14, and 15 February 2011. These three CMEs are named as CME1, CME2, and CME3, respectively. Based on the initial 3D speed and direction of these three CMEs in COR2 FOV, it was evident that they may interact in the interplanetary medium. CME3 was found to be the fastest among all three CMEs and shows a strong deceleration in the COR2 FOV because of the preceding CME2 which acted as a barrier. \cc{Mishra \& Srivastava}  (\cc{2014}) investigated the kinematics of the CMEs in the heliosphere using stereoscopic methods. They noted that a collision between CME3 and CME2 took place around 24 \textit{R}$_\odot$-28 \textit{R}$_\odot$. As CME1 was faint and could not be tracked up to HI2 FOV in \textit{J}-maps, they inferred, based on the extrapolation of distances, that CME2 caught up with CME1 between 138 \textit{R}$_\odot$ to 157 \textit{R}$_\odot$. The kinematics of these three CMEs before and after their interaction is shown in Figure~\cc{\ref{kin_int_CMEsFeb}}. These CMEs were also studied in detail by \cc{Mari\v{c}i\'{c} \etal} (\cc{2014}) using single spacecraft reconstruction methods.

The study of \cc{Mishra \& Srivastava}  (\cc{2014}) identified signatures of collision between CMEs in the kinematics profiles as exchange in their speed. They, under a head-on collision scenario, analysed momentum and energy exchange during the collision phase of CME2 and CME3. They found that collision was close to elastic, as the coefficient of restitution ($e$) was found to be 0.9. However, in another study of the same CMEs, \cc{Mishra \etal} (\cc{2017}) considered an oblique collision scenario for CME2 and CME3, and  found a coefficient of restitution ($e$) of 1.65. This probably suggests that assumption of head-on collision scenario underestimates the value of the coefficient of restitution.

The \textit{in situ} observations, arrival time and geomagnetic response of interacting CMEs of 2011 February 13-15 (CME1, CME2 and CME3) were also studied in \cc{Mishra \& Srivastava} (\cc{2014}). They identified three CMEs as three distict regions in \textit{in situ} observations near 1 AU. The \textit{in situ} observations revealed that CME2 is overheated $\approx$ 10$^{6}$ K, perhaps because it is squeezed between CME1 and CME3. CME2, showing a high speed at the front and low speed at its trailing edge, reveal a signature of fast expansion which was interpreted possibly due to magnetic reconnection at the CME's front edge (\cc{Mari\v{c}i\'{c} \etal 2014}). Such signatures of compression and heating due to CME-CME interaction and passage of CME driven shock through the preceding CME have also been reported in earlier studies (\cc{Lugaz \etal 2005}; \cc{Liu \etal 2012}; \cc{Temmer \etal 2012}). \textit{In situ} data also revealed a smaller spatial scale of CME1 and CME2 than CME3. This is possibly due to compression of preceding CMEs by the following CME or shock. There are also other studies on CME-CME interaction which have shown that interacting CMEs appear as complex ejecta in \textit{in situ} observation and each interacting CME may not be identified as a separate entity (\cc{Liu \etal 2014}; \cc{Lugaz \& Farrugia 2014}; \cc{Mishra \& Srivastava 2014}; \cc{Mishra \etal 2015a},\cc{b}).

\cc{Mishra \etal} (\cc{2016}, \cc{2017}) took into account the propagation and expansion speeds, impact direction, and angular size as well as the masses of the CMEs to understand the CME-CME interaction. They examined for the first time the nature of collision of eight cases of interacting CMEs by carrying out the analysis in 3D scenarios. Among the 8 cases, they showed that the nature of collisions was perfectly inelastic for two cases, inelastic for two cases, elastic for one case, and super-elastic for three cases. The study established that the crucial pre-collision parameters of the CMEs responsible for increasing the probability of a super-elastic collision are, in descending order of priority, their lower approaching speed, expansion speed of the following CME higher than the preceding one, and a longer duration of the collision phase. This important finding is in agreement with the simulation studies (\cc{Shen \etal 2016}). Therefore, it is worth to investigate further the nature of the collision and the processes responsible for magnetic and thermal energy conversion to kinetic energy to make a collision super-elastic.

The observational studies on collision dynamics suffer from uncertainties due to adopted boundary for the start and end of the collision phase (\cc{Mishra \etal 2016}, \cc{2017}). This is because of  difficulty in defining the start of collision as the following CME starts to decelerate (due to its interaction with preceding CME) and preceding CME starts to accelerate before (most possibly due to shock driven by following CME) they both are actually observed to merge. Hence, different timing and large time-interval of acceleration of one CME and deceleration of the other, prevent to pinpoint the exact start and end of the collision phase. Furthermore, the total mass of CMEs is used to study their collision dynamics, but as the CME is not a solid body therefore its total mass is not expected to participate in the collision. Keeping in mind the limitations on the study of CME-CME interaction, further work is required to understand the CME-CME interaction by incorporating various plasma processes.

The geomagnetic response of interacting CMEs has also been investigated extensively in several studies (\cc{Farrugia \etal 2006}; \cc{Mishra \& Srivastava}  (\cc{2014}); \cc{Mishra \etal 2015a}; \cc{Lugaz \& Farrugia 2014}). The study of \cc{Mishra \& Srivastava}  (\cc{2014}) does not favor the possibility of strengthening the geomagnetic response as a consequence of the arrival of two or more interacting CMEs at Earth. However, in another study (\cc{Mishra \etal 2015a}), the interaction region (IR), formed due to the collision between two CMEs, is associated with intensified plasma and magnetic field parameters which were responsible for major geomagnetic activity. The \textit{in situ} measurements of interacting CMEs near 1 AU shows that they are accelerated or decelerated during the interaction, compressed and heated.

The arrival times of interacting CMEs at Earth were also estimated based on HIs observations (\cc{Mishra \etal 2015a}). From the arrival time estimates, it was noticed that arrival time estimation of interacting CMEs improves by few (up to 10) hr when the post-collision speeds are used instead of pre-collision speeds. The estimated post-collision kinematics of interacting CMEs is crucial to be combined with drag based model to improve the arrival time estimation of the interacting CMEs. Thus, several studies in the \textit{STEREO} era reveal that tracking of CMEs up to large heliospheric distances is necessary for better understanding the CME-CME interaction and the prediction of their arrival time at the Earth. Thus, CMEs cannot be treated as completely isolated magnetized plasma blobs, especially when they are launched in quick succession. Each preceding CME offers a different background medium to the following CMEs which should also be taken into account while studying the propagation of CMEs.  From the survey of literature (\cc{Webb \& Howard 2012}; \cc{Harrison \etal 2017}), it is obvious that the prediction of arrival time of CMEs, especially interacting CMEs, and association between remote and \textit{in situ} observations of CMEs, are challenging even in the era of \textit{STEREO} where CMEs can be imaged continuously from near the Sun to near the Earth.

\section{Summary and Future Directions}

The speed, direction, mass and morphology of a CME at a particular location in the heliosphere can be studied by analyzing remote sensing observations while the temperature, speed, density, magnetic field, composition and charge states of CME plasma/solar wind can be measured from \textit{in situ} observations. By the time a CME reaches the spacecraft hosting the instrumentation for \textit{in situ} measurements, it has already evolved and therefore the plasma parameters are different than measured remotely  (\cc{Crooker \& Horbury 2006}). However, if the physics of evolution of a CME is known, then its properties estimated remotely can be extrapolated up to the location of \textit{in situ} spacecraft to make a comparison between both sets of observations with reasonable accuracy. In the absence of a complete understanding of the true nature of the evolution of CMEs, it is often difficult to predict their arrival time to Earth based on their initial characteristics estimated from remote sensing observations made when still near the Sun. The CME characteristics estimated from remote observations suffer from the line of sight integration and projection effects while CME's plasma parameters can be measured along a specific trajectory through the CME by the \textit{in situ} spacecraft (\cc{Webb \& Howard 2012}). The uncertainties in the morphology and kinematics of CMEs due to projection effects, in white light images from a single viewpoint, can be resolved by implementing appropriate 3D reconstruction techniques on the CME images from multiple viewpoints near and far from the Sun.  The continuous spatial coverage of CMEs from the Sun to 1 AU distance was possible by the imaging instruments onboard twin \textit{STEREO} spacecraft. One can unambiguously track a CME continuously from its liftoff in the inner corona to almost the Earth by constructing J-maps (i.e., time-elongation maps) from \textit{STEREO}/SECCHI images  (\cc{Davies \etal 2009}; \cc{M\"{o}stl \etal 2009}). In the present review, a number of benefits of imaging a CME in the heliosphere from the off-Sun-Earth line have been discussed.

Already before the \textit{STEREO} era, it was inferred that CMEs accelerate or decelerate till they obtain the speed of ambient solar wind medium. However, the analysis of several Earth-directed CMEs using the SECCHI/HI observations have helped witnessing such changes in the CME kinematics. Most of the studies in the \textit{STEREO} era, have shown that determining the 3D speed of CMEs near the Sun and assuming that it remains constant for the remaining distance, i.e., up to 1 AU, is not sufficient to accurately predict the arrival time at Earth of the majority of CMEs  (\cc{Harrison \etal 2012}; \cc{Shen \etal 2012b}; \cc{Temmer \etal 2014}; \cc{Mishra \etal 2015a}, \cc{2017}). This is true especially for a fast speed CME traveling in the slow solar wind environment or a slow speed CME traveling in the high speed stream. It is shown that the estimated 3D kinematics of CMEs used as inputs in drag based model (DBM), improve the arrival time prediction of the CMEs at 1 AU. Thus, the role of drag forces, in the dynamics of CMEs, is effective farther out (few tens of solar radii) from the Sun. The studies have also shown that a CME may undergo non-radial longitudinal motion even far from the Sun, specially in the case of CME-CME interaction (\cc{Lugaz \etal 2012}).

The interaction and/or collision of one CME with another CME has been thoroughly investigated in the \textit{STEREO} era (\cc{Harrison \etal 2012}; \cc{Shen \etal 2012b}; \cc{Temmer \etal 2014}; \cc{Mishra \etal 2015a}, \cc{2017}). The studies have concluded that there is a significant exchange of momentum and kinetic energy during the collision of the CMEs. Therefore, post-collision kinematics of CMEs must be used for their improved arrival time prediction at 1 AU. It is also studied that collision/interaction of CMEs have significant effects on the magnetic and plasma parameters of both preceding and following CMEs. The formation of interaction region (IR) at the interface of interacting CMEs is found to be responsible for major geomagnetic activity (\cc{Mishra \etal 2015a}). Therefore, the geomagnetic disturbances due to interacting CMEs need to account for the changes in CMEs parameters resulting from their interaction. Interacting CMEs can be commonly identified in \textit{in situ} observations at 1 AU in the form of multiple-MC events where individual CME can be distinguished, or they appear as a complex ejecta where some of the characteristics of CMEs are lost with shocks propagating inside a previous CME. These structures have different ways to interact with Earth's magnetosphere to give intense geomagnetic storms as compared to an isolated CME. The \textit{in situ} observations at 1 AU shows only the plasma parameters after the CME-CME interaction but not during the collision/interaction duration. The \textit{in situ} data from \textit{Solar Orbiter} and \textit{Parker Solar Probe} can provide the opportunities for measuring the plasma parameters during the interaction process also.

Although HIs have provided the potential to improve the space weather forecasting, some CMEs become too faint to be tracked unambiguously and they impose difficulties in reliably predicting their propagation direction, arrival time, and speed at 1 AU.  Furthermore, the specific assumptions in some of the currently used reconstruction methods compromise the estimates of the complex evolution of the kinematics and morphology of CMEs. Such a complex heliospheric evolution of CMEs is due to their interactions with the ambient slow/fast solar wind, CIRs, and other CMEs which result in errors in the predicted arrival time of CMEs at Earth (\cc{Gopalswamy \etal 2009}; \cc{Lugaz \& Farrugia 2014}; \cc{Mishra \etal 2015a}). Thus, it is important to understand the conditions in the background solar wind and the level of preconditioning of interplanetary medium due to another large-scale solar wind structure, for accurate arrival time prediction of CMEs.

Given the limited success in arrival time prediction of CMEs, it is required to make several studies in this direction (\cc{Harrison \etal 2009}, \cc{Mishra \etal 2014}; \cc{Harrison \etal 2017}). In this regard, it can be advantageous to compare the \textit{J}-maps derived from the observations with the synthetic \textit{J}-maps outputs from the MHD models. Using this comparison of real and synthetic \textit{J}-maps, one can identify the difference in CME evolution in simulation and can eventually correct the model results. Morever, the background solar wind simulated by the models using near-Sun conditions should be refined under the monitoring of other large scale heliospheric structures away from the Sun. It is important to point out that the model-run may provide several solutions under the inputs of different CME and background solar wind parameters. The large number (i.e., spread) in the solutions can be reduced to a small number by comparing it with real \textit{J}-maps. A small number of solutions would then be suitable for predicting a reasonably accurate range of CME arrival times (\cc{Harrison \etal 2017}). Thus, the heliospheric observations have the potential to contribute to operational space weather services.

The \textit{STEREO} era has provided opportunity to understand the association between remotely observed CME structures and \textit{in situ} observations. However, the prediction of negative B$_{z}$ at Earth is most important for predicting the occurrence of geomagnetic storms (\cc{Gonzalez \etal 1989}; \cc{Srivastava \& Venkatakrishnan 2002}). Determination of the negative B$_{z}$ component in the CMEs by exploiting the remote  sensing observations is far from reality. By examining the neutral line in the source region of a CME, one can attempt to guess the inclination of the flux rope, expected direction of rotation and the portion of flux rope where negative B$_{z}$ may occur (\cc{Yurchyshyn \etal 2005}). Our understanding of the flux rope structure of a CME is very limited and it is still debated whether such flux ropes are formed during the eruption or exist before the eruption (\cc{Chen 2011}). 
It is noted that although CME propagation and arrival time are the focus of several studies for long using MHD models and observations, however, the crucial things for space weather prediction for magnetic storms are the direction and intensity of the magnetic field in both the ICMEs and upstream sheath. There are models used to estimate the magnetic field inside the ICMEs arriving at 1 AU; still, such models have not been independently and objectively tested for predictive purposes. There are also quite promising studies using machine learning algorithms to predict geomagnetic storms at 1 AU (\cc{Pricopi \etal 2022}). But again, one should properly test the reliability of such approaches. Thus, further work is required to understand the key issues responsible for space weather near Earth.

The successful exploitation of heliospheric imagery can revolutionize our understanding of the evolution of CMEs. This have made researchers to include instruments similar to HIs on other space missions such as SoloHI (\cc{Howard \etal 2020}) on the \textit{Solar Orbiter (SO)} (\cc{M\"{u}ller \etal 2020}) and the Wide-field Imager for Solar PRobe (WISPR) (\cc{Vourlidas \etal 2016}) on the \textit{Parker Solar Probe (PSP)} (\cc{Fox \etal 2016}). The orbital motion of \textit{STEREO} has allowed heliospheric imaging from different heliospheric locations including their passage through the L4 and L5 Lagrangian points. With the progress of the \textit{STEREO} mission, the separation between the twin spacecraft reached to 180$^{\circ}$) around the beginning of year 2011, and further they continued increasing their separation from the Sun-Earth line. These locations of \textit{STEREO}, behind the Sun from Earth's viewpoint, are less suitable for the heliospheric imaging of the Earth-directed CMEs. This is most importantly because an Earth-directed CME tend to lie outside the Thomson sphere and are poorly visible from these \textit{STEREO} locations behind the Sun (\cc{Howard \& DeForest 2012}). Also, the communications with \textit{STEREO-B} were lost around October 2014 which were re-established only in August 2016, and it has now been out of contact since September 2016. Although \textit{STEREO-A} continues to operate normally, the loss of \textit{STEREO-B} somewhat limited the operational potential of \textit{STEREO} mission. The higher signal-to-noise imaging observations from heliospheric imagers onboard \textit{PSP} and \textit{SO} allow now imaging of the inner and darker cavities instead of traditional tracking of bright fronts of CMEs/ICMEs using \textit{STEREO}. The traditional heliospheric tracking of ICMEs in white-light imaging observations identifies the enhanced density structure, which is the compressed sheath region, whereas the \textit{in situ} observations accurately identify the density-depleted structure with the enhanced magnetic field, which is the magnetic ejecta or the flux-rope. The flux rope structure of ICMEs appears as the darker cavity in the white-light imaging observations and therefore, the tracking of the darker cavity can enable to more accurately connect the remote sensing observations of ICMEs with their \textit{in situ} observations (\cc{Hess \etal 2020}; \cc{Howard \etal 2020}; \cc{Rouillard \etal 2020}). Furthermore, observations from these new missions may change a few of the assumptions on the radial evolution of plasma parameters (i.e., density, magnetic field, etc.) in CMEs and solar wind, and therefore it has the potential to refine the existing propagation models (empirical or MHD) of CMEs.

The \textit{STEREO}/HI observations almost always have been used in conjunction with \textit{SOHO}/LASCO observations to study the heliospheric evolution of CMEs. A similar approach should also be taken for the observations of \textit{PSP}/WISPR, SoloHI and of the Metis coronagraph onboard \textit{SO}. We expect that future studies will provide further insights by coordinated science campaigns with multiple space missions like \textit{SOHO}, \textit{STEREO}, \textit{SDO}, \textit{PSP}, \textit{SO} and so on. The observations of a large number of CMEs at different heliocentric distances from the Sun by \textit{PSP} and \textit{SO} missions, may help in validating the models of CME magnetic field forecasting. In a recent study modeling the evolution of ICMEs, the researchers have observed the plasma parameters of the ICMEs by widely separated five \textit{in situ} spacecraft, \textit{SO}, \textit{BepiColombo}, \textit{PSP}, \textit{Wind}, and \textit{STEREO-A}, in connection with the remote observations of the same ICMEs by coronagraph and heliospheric imager onboard \textit{STEREO-A}/SECCHI and \textit{SOHO}/LASCO (\cc{M{\"o}stl \etal 2022}). Such studies on several cases, possible during the maximum phase of solar cycle 25, can improve understanding of the interplanetary evolution of ICMEs, their magnetic structure, global shape of their flux ropes, and shocks. Further, the \textit{in-situ} monitoring of ICMEs at Venus orbit combined with empirical and/or propagation models may provide early predictions of Earth-bound CMEs. As \textit{PSP} is measuring the regions where the solar wind gets accelerated, it is important to study fast solar wind flow from coronal holes and its impacts on the CMEs evolution. \textit{SO} progressively inclined orbit over the ecliptic will provide new insight onto the polar regions of the Sun and is expected to improve our understanding of the solar wind from coronal holes in the polar regions. Future studies should focus on verifying and evaluating background solar wind models to improve the inputs for CME propagation models. The studies of CMEs and solar wind evolution will be further augmented by the anticipated launches of the ESA \textit{Proba-3} (\cc{Shestov \etal 2021}) satellites in 2022 and the \textit{Polarimeter to UNify the Corona and heliosphere} \textit{(PUNCH)} (\cc{DeForest \etal 2020}) in 2023. The observations from such upcoming missions will be highly complementary to both WISPR  onboard \textit{PSP} and SoloHI onboard \textit{SO}.

In light of the discussions made above, it is clear that little progress has been made in accurately estimating the arrival time, arrival speed, size, mass, magnetic field configuration, and field strength of a particular CME at a location in the heliosphere. This is because various observational and modeling limitations partially prevent a more accurate determination of the CME arrival time. Also, our limited understanding of the physics of solar wind in the inner heliosphere and the immense size of the physical system we are dealing with, further prevent accuracy in the current trends of research (\cc{Harrison \etal 2017}). There has been good progress in understanding the dynamics of CMEs under the influence of high speed from coronal holes, other CMEs, and the ambient pre-conditioned medium, but yet we are far from the complete understanding (\cc{Manchester \etal 2017}). We note that the energy budget of CMEs has only been studied for a handful of cases within a few solar radii from the Sun. Also, we do not have a good understanding of the shape, size, and structure of CME’s front and shock which also poses challenges for estimating the accurate arrival time of CMEs. It is still difficult to reliably track the evolution of different CME structures, particularly the magnetic flux rope (MFR). Furthermore, the limited knowledge of the physical parameters in the near corona hinders robust modeling of the initial stages of CME propagation and shock evolution. From the space weather perspective, the magnetic properties of the CMEs are often not reliably estimated near the Sun. Also, the heliospheric evolution of the CME magnetic structure (rotation, compression, deflection) and the erosion of the magnetic field due to reconnection with ambient solar wind magnetic fields pose considerable difficulty in predicting both the magnitude and geometry of the CME magnetic field at 1 AU (\cc{Wang \etal 2018}). Although there have been considerable developments in heliospheric imaging, it has remained extremely difficult to predict the duration of CMEs impacts and their momentum at the Earth.

Despite several limitations to HIs, it seems that there is no better substitute for imaging the vast and crucial distance gap between the Sun and Earth. This is because it is unlikely that MHD modeling would realistically predict the conditions of the ambient medium for estimating the complex evolution of the CMEs. Therefore, monitoring the CMEs during their continuous journey from the Sun to Earth has the potential to reveal the physics of evolution of the CMEs. In the future, we expect that a stationary spacecraft outside the Sun-Earth line (e.g., a space weather mission to the L5 point of the Sun-Earth system is being developed by ESA to be launched in 2027), continuously imaging the heliosphere from a stable platform, can overcome some of the limitations suffered by the \textit{STEREO}. The spacecraft at L4/L5 Lagrange points giving necessary side views of the Sun will observe Earth-directed CMEs with low projection effects. Such spacecraft providing real-time telemetry of good quality data can play a crucial role in achieving a credible space weather prediction. Also, the proposed polar missions, \textit{Solaris Solar Polar Mission}, if approved, would provide unprecedented observations to improve the understanding of magnetic field connectivity and coupling processes between open and closed magnetic field structures in the heliosphere. At present,  the valuable heliospheric observations from recent missions are waiting to be explored extensively. The analysis of these unprecedented observations has the potential to improve our understanding of CME propagation and the performance of space weather prediction tools and models.

\section*{Acknowledgements}
We thank the editorial board of the Journal of Astrophysics and Astronomy (JoAA) for inviting WM to write this review article. We thank the publisher AAS to grant permission to reproduce some figures from The Astrophysical Journal, and Springer for permitting us to reproduce some figures from the journals of Solar Physics and Space Science Review. We also thank Nandita Srivastava (USO, India) for helpful suggestions. The authors are thankful to the referee for his/her comments which have improved the manuscript.


\vspace{-1em}
\raggedright

\end{document}